\documentstyle[amssymb,prl,twocolumn,epsf,aps]{revtex}

\begin{document}

\twocolumn[\hsize\textwidth\columnwidth\hsize\csname @twocolumnfalse\endcsname
\title{Quantum versus Classical Domains for Teleportation
with Continuous Variables}

\author{Samuel\ L.\ Braunstein$^{\dagger}$, Christopher A.\
Fuchs$^{\ddagger}$, H.\ J.\ Kimble$^{\ast}$, and P. van
Loock$^{\dagger}$}

\address{$^\dagger$Informatics, Bangor University, Bangor LL57 1UT, UK}
\address{$^\ddagger$ Bell Labs, Lucent Technologies,
600-700 Mountain Ave., Room 2C-420,
Murray Hill, NJ 07974, USA}

\address{$^\ast$Norman Bridge Laboratory of Physics 12-33,
California Institute of Technology,
Pasadena, California 91125, USA}

\date{\today}
\maketitle

\begin{abstract}
Fidelity $F_{classical}=\frac{1}{2}$ has been established as
setting the boundary between classical and quantum domains in the
teleportation of coherent states of the electromagnetic field (S. L.
Braunstein, C. A. Fuchs, and H. J. Kimble, J. Mod. Opt. {\bf 47},
267 (2000)). Two recent papers by P. Grangier and F. Grosshans ({\tt
quant-ph/0009079} and {\tt quant-ph/0010107}) introduce alternate
criteria for setting this boundary and as a result claim that the appropriate
 boundary should be
$F=\frac{2}{3}$. Although larger fidelities would lead to enhanced
teleportation capabilities, we show that the new conditions of Grangier
and Grosshans are largely unrelated to the questions of entanglement
and Bell-inequality violations that they take to be their primary
concern. With regard to the quantum-classical
boundary, we demonstrate that fidelity $F_{classical}=\frac{1}{2}$
remains the appropriate point of demarcation. The claims of Grangier
and Grosshans to the contrary are simply wrong, as we show by an
analysis of the conditions for nonseparability (that complements our
earlier treatment) and by explicit examples of Bell-inequality
violations.
\end{abstract}
\vspace{3ex}
]

\section{Introduction}

As proposed by Bennett {\it et al.}~\cite{Bennett93}, the protocol for
achieving quantum teleportation is the following. Alice is to transfer an
unknown quantum state $|\psi \rangle $ to Bob, using as the sole resources
some previously shared {\it quantum entanglement} and a {\it classical
channel\/} capable of communicating measurement results. Physical transport
of $|\psi \rangle $ from Alice to Bob is excluded at the outset. Ideal
teleportation occurs when the state $|\psi \rangle $ enters Alice's sending
station and the {\it same} state $|\psi \rangle $ emerges from Bob's
receiving station.

Of course, in actual experiments\cite
{Bouwmeester97,Boschi98,Furusawa98,Nielsen98}, the ideal case is {\it %
unattainable\/} as a matter of principle. The question of operational
criteria for gauging success in an experimental setting, therefore, cannot
be avoided. We have proposed previously that a minimal set of conditions for
claiming success in the laboratory are the following\cite{Fuchs00}.

\begin{enumerate}
\item  An unknown quantum state (supplied by a third party Victor) is input
physically into Alice's station from an outside source.

\item  The ``recreation'' of this quantum state emerges from Bob's receiving
terminal available for Victor's independent examination.

\item  There should be a quantitative measure for the quality of the
teleportation and based upon this measure, it should be clear that shared
entanglement enables the output state to be ``closer'' to the input state
than could have been achieved if Alice and Bob had utilized a classical
communication channel alone.
\end{enumerate}

In Ref.\cite{Fuchs00}, it was shown that the fidelity $F$ between input and
output states is an appropriate measure of the degree of similarity in
Criterion $3$. For an input state $|\psi _{in}\rangle $ and output state
described by the density operator $\hat{\rho}_{out}$, the fidelity is given
by\cite{Schumacher95}
\begin{equation}
F=\langle \psi _{in}|\hat{\rho}_{out}|\psi _{in}\rangle \text{ .}
\label{fidelity}
\end{equation}
To date only the experiment of Furusawa {\it et al.}\cite{Furusawa98} has
achieved unconditional experimental teleportation as defined by the three
criteria above\cite{Fuchs00,Braunstein98a,Braunstein98b}. This experiment
was carried out in the setting of continuous quantum variables with input
states $|\psi _{in}\rangle $ consisting of coherent states of the
electromagnetic field, with an observed fidelity $F_{\exp }=0.58\pm 0.02$
having been attained. This benchmark is significant because it can be
demonstrated\cite{Furusawa98,Fuchs00} that quantum entanglement is the
critical ingredient in achieving an average fidelity greater than $%
F_{classical}=\frac{1}{2}$ {\it when\/} the input is an absolutely random
coherent state\cite{explainf}.

Against this backdrop, Grangier and Grosshans\cite{Grangier00a,Grangier00b}
have recently suggested that the appropriate boundary between the classical
and quantum domains in the teleportation of coherent states should be a
fidelity $F=\frac{2}{3}$. Their principal concern is the distinction between
``entanglement'' and ``non-separability,'' where by the latter term, they
mean ``the physical properties associated with non-locality and the
violation of Bell's inequalities (BI).''\footnote{%
Since the terms ``entanglement'' and ``nonseparability'' are used
interchangeably in the quantum information community, we will treat them as
synonyms to eliminate further confusion. We will refer to violations of
Bell's inequalities explicitly whenever a distinction must be made between
entanglement and local realism {\it per se}. The only exceptions will be
when we quote directly from Grangier and Grosshans\cite
{Grangier00a,Grangier00b}.} They claim that ``due to imperfect
transmissions, ... it becomes possible to violate the classical boundary
{\it (i.e., }$F=\frac{1}{2}${\it )} of teleportation without any violation
of BI.''\cite{Grangier00a} However, rather than addressing the issue in a
direct manner, they then propose the violation of a certain
``Heisenberg-type inequality (HI)'' as ``a more effective -- and in some
sense `necessary' -- way to characterize shared entanglement.'' It is this
that leads to their condition $F>\frac{2}{3}$ as being necessary for the
declaration of successful teleportation. In support of this threshold, they
further relate their criterion based on the HI to ones previously introduced
in the quantum nondemolition measurement (QND) literature. Finally, in Ref.
\cite{Grangier00b}, Grangier and Grosshans find that $F>\frac{2}{3}$ is also
required by a criterion they introduce having to do with a certain notion of
reliable ``information exchange''\cite{Grangier00b}.

The purpose of the present paper is to demonstrate that the conclusions of
Grangier and Grosshans concerning the proposed quantum-classical boundary $F=%
\frac{2}{3}$ are unwarranted and, by explicit counter example, incorrect.
Our approach will be to investigate questions of nonseparability and
violations of Bell inequalities for the particular entangled state employed
in the teleportation protocol of Ref.\cite{Braunstein98c}. Of significant
interest will be the case with losses, so that the relevant quantum states
will be mixed quantum states. Our analysis supports the following
conclusions.

\begin{enumerate}
\item  Although the argument of Grangier and Grosshans is claimed to be
based upon ``EPR non-separability of the entanglement resource''\cite
{Grangier00b} [by which they mean a potential violation of a BI], they offer
no quantitative connection (by constructive proof or otherwise) between the
criteria they introduce (including the threshold $F=\frac{2}{3}$)\cite
{Grangier00a,Grangier00b} and the actual violation of any Bell inequality.
Nothing in their analysis provides a warranty that $F>\frac{2}{3}$ would
preclude a description in terms of a local hidden-variables theory. They
offer only the suggestion that ``$F>\frac{2}{3}$ would be much safer''\cite
{Grangier00a}.

\item  By application of the work of Duan {\it et al.}\cite{duan00}, Simon
\cite{simon00}, and Tan\cite{tan99}, we investigate the question of
entanglement. We show that the states employed in the experiment of Ref.\cite
{Furusawa98} are nonseparable, as was operationally confirmed in the
experiment. Moreover, we study the issue of nonseparability for mixed states
over a broad range in the degree of squeezing for the initial EPR state, in
the overall system loss, and in the presence of thermal noise. This analysis
reveals that EPR mixed states that are nonseparable do indeed lead to a
fidelity of $F>F_{classical}=\frac{1}{2}$ for the teleportation of coherent
states. Hence, in keeping with Criterion $3$ above, the threshold fidelity
for employing entanglement as a quantum resource is precisely the same as
was deduced in the previous analysis of Ref.\cite{Fuchs00}. Within the
setting of Quantum Optics, this threshold coincides with the standard
benchmark for manifestly quantum or nonclassical behavior, namely that the
Glauber-Sudarshan phase-space function becomes nonpositive-definite, here
for any bipartite nonseparable state\cite{simon00-gs}. By contrast, the
value $F=\frac{2}{3}$ championed by Grangier and Grosshans is essentially
unrelated to the threshold for entanglement (nonseparability) in the
teleportation protocol, as well as to the boundary for the nonclassical
character of the EPR state.

\item  By application of the work of Banaszek and Wodkiewicz\cite
{b-k98,b-k99}, we explore the possibility of violations of Bell inequalities
for the EPR (mixed) states employed in the teleportation of continuous
quantum-variables states. We find direct violations of a CHSH inequality\cite
{c-s78} over a large domain. Significant relative to the claims of Grangier
and Grosshans is a regime both of entanglement (nonseparability) and of
violation of a CHSH inequality for which the teleportation fidelity $F<\frac{%
2}{3}$ and for which the criterion of the Heisenberg inequalities of Ref.
\cite{Grangier00a} fails. Hence, teleportation with $\frac{1}{2}<F<\frac{2}{3%
}$ is possible with EPR (mixed) states which do not admit a local hidden
variables description. In contradistinction to the claim of Grangier and
Grosshans, $F>\frac{2}{3}$ does not provide a relevant criterion for
delineating the quantum and classical domains with respect to violations of
Bell's inequalities.

\item  By adopting a protocol analogous to that employed in {\it all\/}
previous experimental demonstrations of violations of Bell's inequalities
\cite{Scully97,Kwiat94,Fry95}, scaled correlation functions can be
introduced for continuous quantum variables. In terms of these scaled
correlations, the EPR mixed state used for teleportation violates a
generalized version of the CHSH inequality, though non-ideal detector
efficiencies require a ``fair sampling'' assumption for this. These
violations set in for $F>\frac{1}{2}$ and have recently been observed in a
setting of low detection efficiency\cite{Kuzmich00}. This experimental
verification of a violation of a CHSH inequality (with a fair sampling
assumption) again refutes the purported significance of the threshold $F=%
\frac{2}{3}$ promoted by Grangier and Grosshans.
\end{enumerate}

Overall, we find no support for the claims of Grangier and Grosshans giving
special significance to the threshold fidelity $F=\frac{2}{3}$ in connection
to issues of separability and Bell-inequality violations. Instead, as we
will show, it is actually the value $F_{classical}=\frac{1}{2}$ that heralds
entrance into the quantum domain with respect to the very same issues. Their
claims based upon a Heisenberg-type inequality and a criterion for
``information exchange'' are essentially unrelated to the issue of a
quantum-classical boundary.

All this is not to say that teleportation of coherent states with increasing
degrees of fidelity beyond $F_{classical}=\frac{1}{2}$ to $F>\frac{2}{3}$ is
not without significance. In fact, as tasks of ever increasing complexity
are to be accomplished, there will be corresponding requirements to improve
the fidelity of teleportation yet further. Moreover, there are clearly
diverse quantum states other than coherent states that one might desire to
teleport, including squeezed states, quantum superpositions, entangled
states \cite{tan99}, and so on. The connection between the ``intricacy'' of
such states and the requisite resources for achieving high fidelity
teleportation has been discussed in Ref.\cite{Braunstein98c}, including the
example of the superposition of two coherent states,
\begin{equation}
|\alpha \rangle +|-\alpha \rangle \text{ ,}
\end{equation}
which for $|\alpha |\gg 1$ requires an EPR\ state with an extreme degree of
quantum correlation.

Similiarly, Heisenberg-type inequalities are in fact quite important for the
inference of the properties of a {\it system} given the outcomes of
measurements made on a {\it meter} following a {\it system-meter }%
interaction. Such quantities are gainfully employed in Quantum Optics in
many settings, including realizations of the original EPR {\it gedanken}
experiment\cite{EPR,Ou92,reid89} and of back-action evading measurement and
quantum non-demolition detection\cite{Grangier98}.

Our only point is that the claim of Grangier and Grosshans that $F=\frac{2}{3%
}$ is required for the ``successful quantum teleportation of a coherent
state''\cite{Grangier00b} is incorrect. They simply offer no quantitative
analysis directly relevant to either entanglement or Bell-inequality
violation issues. In contrast, the prior treatment of Ref.\cite{Fuchs00}
demonstrates that in the absence of shared entanglement between Alice and
Bob, there is an upper limit for the fidelity for the teleportation of
randomly chosen coherent states given by $F_{classical}=\frac{1}{2}$.
Nothing in the work of Grangier and Grosshans calls this analysis into
question.

This, however, leads to something we would like to stress apart from the
details of any particular teleportation criterion. There appears to be a
growing confusion in the community that equates quantum teleportation
experiments with fundamental tests of quantum mechanics. The purpose of such
tests is generally to compare quantum mechanics to other potential theories,
such as locally realistic hidden-variable theories\cite
{Grangier00a,CavesPrivate,WodkiewiczPrivate}. In our view, experiments in
teleportation have nothing to do with this. They instead represent
investigations {\it within\/} quantum mechanics, demonstrating only that a
particular task can be accomplished with the resource of quantum
entanglement and cannot be accomplished without it. This means that
violations of Bell's inequalities are largely irrelevant as far as the
original proposal of Bennett {\it et al.}\cite{Bennett93} is concerned, as
well as for experimental implementations of that protocol. In a theory which
allows states to be cloned, there would be no need to discuss teleportation
at all -- unknown states could be cloned and transmitted with fidelity
arbitrarily close to one.

These comments notwithstanding, Grangier and Gross\-hans did nevertheless
attempt to link the idea of Bell-inequality violations with the fidelity of
teleportation. It is to the details of that linkage that we now turn. The
remainder of the paper is organized as follows. In Section II, we extend the
prior work of Ref.\cite{Fuchs00} to a direct treatment of the consequences
of shared entanglement between Alice and Bob, beginning with an explicit
model for the mixed EPR states used for teleportation of continuous quantum
variables. In Section III we review the criteria Grangier and Grosshans
introduced in preparation for showing their inappropriateness as tools for
the questions at hand. In Section IV, we demonstrate explicitly the
relationship between entanglement and fidelity, and find the same threshold $%
F_{classical}=\frac{1}{2}$ as in our prior analysis\cite{Fuchs00}. The value
$F=\frac{2}{3}$ is shown to have no particular distinction in this context.
In Sections V and VI, we further explore the role of entanglement with
regard to violations of a CHSH inequality and provide a quantitative
boundary for such violations. Again, $F_{classical}=\frac{1}{2}$ appears as
the point of entry into the quantum domain, with the point $F=\frac{2}{3}$
having no particular distinction. Our conclusions are collected in Section
VII. Of particular significance, we point out that the teleportation
experiment of Ref.\cite{Furusawa98} did indeed cross from the classical to
the quantum domain, just as advertised previously.

\section{The EPR State}

The teleportation protocol we consider is that of Braunstein and Kimble\cite
{Braunstein98c}, for which the relevant entangled state is the so-called
two-mode squeezed state. This state is given explicitly in terms of a
Fock-state expansion for two-modes $(1,2)$ by\cite{walls,vanEnk99}
\begin{equation}
|EPR\rangle _{{\rm \scriptscriptstyle}1,2}=\frac{1}{\cosh r}%
\sum_{n=0}^{\infty }(\tanh r)^{n}|n\rangle _{{\rm \scriptscriptstyle}%
1}|n\rangle _{{\rm \scriptscriptstyle}2}\;,  \label{eprstate}
\end{equation}
where $r$ measures the amount of squeezing required to produce the entangled
state. Note that for simplicity we consider the case of two single modes for
the electromagnetic field; the extension to the multimode case for fields of
finite bandwidth can be found in Ref.\cite{vanLoock00}.

The pure state of Eq.~(\ref{eprstate}) can be equivalently described by the
corresponding Wigner distribution $W_{{\rm EPR}}$ over the two modes $(1,2)$%
,
\begin{eqnarray}
&&W_{{\rm EPR}}(x_{1},p_{1};x_{2},p_{2})  \label{wepr} \\
&=&{\frac{4}{\pi ^{2}}}\frac{1}{\sigma _{+}^{2}\sigma _{-}^{2}}\,\exp \Big(%
-[(x_{1}+x_{2})^{2}+(p_{1}-p_{2})^{2}]/\sigma _{+}^{2}  \nonumber \\
&&-[(x_{1}-x_{2})^{2}+(p_{1}+p_{2})^{2}]/\text{ }\sigma _{-}^{2}\Big)\text{ ,%
}  \nonumber
\end{eqnarray}
where $\sigma _{\pm }$ are expressed in terms of the squeezing parameter by
\begin{eqnarray}
\sigma _{+}^{2} &=&e^{+2r},\text{ } \\
\sigma _{-}^{2} &=&e^{-2r}\text{,}  \nonumber
\end{eqnarray}
with $\sigma _{+}^{2}\sigma _{-}^{2}=1$. Here, the canonical variables $%
(x_{j},p_{j})$ are related to the complex field amplitude $\alpha _{j}$ for
mode $j=(1,2)$ by
\begin{equation}
\alpha _{j}=x_{j}+ip_{j}\text{.}
\end{equation}
In the limit of $r\rightarrow \infty $, Eq.~(\ref{wepr}) becomes
\begin{equation}
C\,\delta (x_{1}-x_{2})\,\delta (p_{1}+p_{2})  \label{rlimit}
\end{equation}
which makes a connection to the original EPR state of Einstein, Podolsky,
and Rosen\cite{EPR}.

Of course, $W_{{\rm EPR}}$ as given above is for the ideal, lossless case.
Of particular interest with respect to experiments is the inclusion of
losses, as arise from, for example, finite propagation and detection
efficiencies. Rather than deal with any detailed setup (e.g., as treated in
explicit detail in Ref.\cite{Ou92}) here we adopt a generic model of the
following form. Consider two identical beam splitters each with a
transmission coefficient $\eta$, one for each of the two EPR modes. We take $%
0\leq \eta \leq 1$, with $\eta =1$ for the ideal, lossless case. The input
modes to the beam splitter $1$ are taken to be $(1^{\prime },a^{\prime })$,
while for beam splitter $2$, the modes are labeled by $(2^{\prime
},b^{\prime })$. Here, the modes $(1^{\prime },2^{\prime })$ are assumed to
be in the state specified by the ideal $W_{{\rm EPR}}$ as given in Eq.~(\ref
{wepr}) above, while the modes $(a^{\prime },b^{\prime })$ are taken to be
independent thermal (mixed) states each with Wigner distribution
\begin{equation}
W(x,p)=\frac{1}{\pi (\bar{n}+\frac{1}{2})}\exp \{-(x^{2}+p^{2})/(\bar{n}
+1/2)\}\text{,}
\end{equation}
where $\bar{n}$ is the mean thermal photon number for each of the modes $%
(a^{\prime },b^{\prime })$.

The overall Wigner distribution for the initial set of input modes $%
(1^{\prime },2^{\prime }),(a^{\prime },b^{\prime })$ is then just the
product
\begin{equation}
W_{{\rm EPR}}(x_{1^{\prime }},p_{1^{\prime }};x_{2^{\prime }},p_{2^{\prime
}})W(x_{a^{\prime }},p_{a^{\prime }})W(x_{b^{\prime }},p_{b^{\prime }})\text{
.}
\end{equation}
The standard beam-splitter transformations lead in a straightforward fashion
to the Wigner distribution for the output set of modes $(1,2),(a,b)$, where,
for example,
\begin{eqnarray}
x_{1} &=&\sqrt{\eta }x_{1^{\prime }}-\sqrt{1-\eta }x_{a^{\prime }} \text{,}
\label{bs} \\
x_{a} &=&\sqrt{\eta }x_{a^{\prime }}+\sqrt{1-\eta }x_{1^{\prime }} \text{.}
\nonumber
\end{eqnarray}

We require $W_{{\rm EPR}}^{out}$ for the $(1,2)$ modes alone, which is
obtained by integrating over the $(a,b)$ modes. A straightforward
calculation results in the following distribution for the mixed output
state:
\begin{eqnarray}
&&W_{{\rm EPR}}^{out}(x_{1},p_{1};x_{2},p_{2})  \label{wout} \\
&=&{\frac{4}{\pi ^{2}}}\frac{1}{\bar{\sigma}_{+}^{2}\bar{\sigma}_{-}^{2}}\,
\exp \Big(-[(x_{1}+x_{2})^{2}+(p_{1}-p_{2})^{2}]/\bar{\sigma}_{+}^{2}
\nonumber \\
&&~-[(x_{1}-x_{2})^{2}+(p_{1}+p_{2})^{2}]/\text{ }\bar{\sigma}_{-}^{2}\Big)
\text{,}  \nonumber
\end{eqnarray}
where $\bar{\sigma}_{\pm }$ are given by
\begin{eqnarray}
\bar{\sigma}_{+}^{2} &=&\eta e^{+2r}+(1-\eta )(1+2\bar{n}),\text{ } \\
\bar{\sigma}_{-}^{2} &=&\eta e^{-2r}+(1-\eta )(1+2\bar{n})\text{.}  \nonumber
\end{eqnarray}
Note that $W_{{\rm EPR}}^{out}$ as above follows directly from $W_{{\rm EPR}%
} $ in Eq.~(\ref{wepr}) via the simple replacements $\sigma _{\pm
}\longrightarrow \bar{\sigma}_{\pm }$. Relevant to the discussion of Bell
inequalities in Sections V and VI is the fact that $\bar{\sigma}_{+}^{2}\bar{%
\sigma}_{-}^{2}>1$ for any $r>0$ and $\eta <1$.

\section{The Criteria of Grangier and Grosshans}

The two recent papers of Grangier and Grosshans argue that ``fidelity value
larger than $\frac{2}{3}$ is actually required for successful
teleportation'' \cite{Grangier00a,Grangier00b}. In this section, we
recapitulate the critical elements of their analysis and state their
criteria in the present notation. In subsequent sections we proceed further
with our own analysis of entanglement and possible violations of Bell's
inequalities for the EPR state of Eq.~(\ref{wout}).

Beginning with Ref.\cite{Grangier00a}, Eq.~(21), Grangier and Grosshans
state the following:

\begin{quote}
``As a criteria for non-separability [by which they mean violations of
Bell's inequalities], we will use the EPR\ argument: two different
measurements prepare two different states, in such a way that the product of
conditional variances (with different conditions) violates the Heisenberg
principle.''
\end{quote}

This statement takes a quantitative form in terms of the following
conditional variances expressed in the notation of the preceding section for
EPR\ beams $(1,2)$:
\begin{eqnarray}
V_{x_{i}|x_{j}} &=&\langle \Delta x_{i}^{2}\rangle -\frac{\langle
x_{i}x_{j}\rangle ^{2}}{\langle \Delta x_{j}^{2}\rangle }\text{, }
\label{conditionalvariances} \\
\text{\ }V_{p_{i}|p_{j}} &=&\langle \Delta p_{i}^{2}\rangle -\frac{\langle
p_{i}p_{j}\rangle ^{2}}{\langle \Delta p_{j}^{2}\rangle }\text{.}  \nonumber
\end{eqnarray}
with $(i,j)=(1,2)$ and $i\neq j$. Note that, for example, $V_{x_{2}|x_{1}}$
gives the error in the knowledge of the canonical variable $x_{2}$ based
upon an estimate of $x_{2}$ from a measurement of $x_{1}$, and likewise for
the other conditional variances. These variances were introduced in Refs.
\cite{Ou92,reid89} in connection with an optical realization of the original
{\it gedanken }experiment of Einstein, Podolsky, and Rosen\cite{EPR}. An
apparent violation of the uncertainty principle arises if the product of
inference errors is below the uncertainty product for one beam alone. For
example, $V_{x_{2}|x_{1}}V_{p_{2}|p_{1}}<\frac{1}{16}$ represents such an
apparent violation since $\Delta x_{2}^{2}\Delta p_{2}^{2}\geq \frac{1}{16}$
is demanded by the canonical commutation relation between $x_{2}$ and $p_{2}$%
, with here $\Delta x_{1,2}^{2}=\frac{1}{4}=$ $\Delta p_{1,2}^{2}$ for the
vacuum state\cite{Ou92,reid89}.

Grangier and Grosshans elevate this concept of inference at a distance from
the EPR\ analysis to ``a criteria for non-separability [i.e., violation of
Bell's inequalities].'' Specifically, they state that ``the classical limit
of no apparent violation of HI'' [and hence the domain of local realism] is
determined by the conditions
\begin{equation}
V_{x_{2}|x_{1}}V_{p_{2}|p_{1}}\geq \frac{1}{16}\text{, \ \ and \ \ }%
V_{x_{1}|x_{2}}V_{p_{1}|p_{2}}\geq \frac{1}{16}\text{. }  \label{HI}
\end{equation}
As shown in Refs.\cite{Ou92,reid89} for the states under consideration, the
conditional variances of Eq.~(\ref{conditionalvariances}) are simply related
to the following (unconditional) variances
\begin{eqnarray}
\Delta x_{\mu _{ij}}^{2} &=&\langle (x_{i}-\mu _{ij}x_{j})^{2}\rangle ,
\label{munuvar} \\
\Delta p_{\nu _{ij}}^{2} &=&\langle (p_{i}-\nu _{ij}p_{j})^{2}\rangle \text{.%
}  \nonumber
\end{eqnarray}
If we use a measurement of $x_{j}$ to estimate $x_{i}$, then $\Delta x_{\mu
_{ij}}^{2}$ is the variance of the error when the estimator is chosen to be $%
\mu _{ij}x_{j}$, and likewise for $\Delta p_{\nu _{ij}}^{2}$. For an optimal
estimate, the parameters $(\mu _{ij},\nu _{ij})$ are given by\cite
{Ou92,reid89}
\begin{equation}
\mu _{ij}^{\text{opt}}=\frac{\langle x_{i}x_{j}\rangle }{\langle \Delta
x_{j}^{2}\rangle }\text{, \ \ \ }\nu _{ij}^{\text{opt}}=\text{\ }\frac{%
\langle p_{i}p_{j}\rangle }{\langle \Delta p_{j}^{2}\rangle }\text{,}
\label{munu}
\end{equation}
and in this case,
\begin{equation}
V_{x_{i}|x_{j}}=\Delta x_{\mu _{ij}^{\text{opt}}}^{2}\text{,}\ \text{\ and }%
V_{p_{i}|p_{j}}=\Delta p_{\nu _{ij}^{\text{opt}}}^{2}\text{.}
\end{equation}
The ``non-separability'' condition of Grangier and Grosshans in Eq.~(\ref{HI}
) can then be re-expressed as
\begin{equation}
\Delta x_{\mu _{21}}^{2}\Delta p_{\nu _{21}}^{2}\geq \frac{1}{16}\text{, \ \
and \ \ }\Delta x_{\mu _{12}}^{2}\Delta p_{\nu _{12}}^{2}\geq \frac{1}{16}%
\text{,}  \label{newHI}
\end{equation}
where we assume the optimized choice and drop the superscript `opt'. Again,
Grangier and Grosshans take this condition of \ ``no apparent violation of
HI'' as the operational signature of ``nonseparability criteria''
[violations of Bell inequalities], and hence, by their logic, to delineate
the classical boundary for teleportation\cite{Grangier00a}.

To make apparent the critical elements of the discussion, we next assume
symmetric fluctuations as appropriate to the EPR state of Eq.~(\ref{wout}), $%
\mu _{ij}=\mu _{ji}\equiv \mu $ and $\nu _{ij}=\nu _{ji}\equiv \nu $, with $%
\mu =-\nu $. Note that within the context of our simple model of the losses,
the optimal value of $\mu $ is given by
\begin{equation}
\mu =\frac{\eta \sinh 2r}{(1-\eta )+\eta \cosh 2r}\text{,}  \label{muopt}
\end{equation}
where in the limit $r\gg 1$, $\mu \rightarrow 1$. For this case of symmetric
fluctuations, the HI of Eq. (\ref{newHI}) becomes
\begin{equation}
\Delta x_{\mu }^{2}\Delta p_{\mu }^{2}\geq \frac{1}{16}\text{,}  \label{GGHI}
\end{equation}
where
\begin{eqnarray}
\Delta x_{\mu }^{2} &=&\langle (x_{1}-\mu x_{2})^{2}\rangle =\langle
(x_{2}-\mu x_{1})^{2}\rangle ,  \label{dxdpmu} \\
\Delta p_{\mu }^{2} &=&\langle (p_{1}+\mu p_{2})^{2}\rangle =\langle
(p_{2}+\mu p_{1})^{2}\rangle \text{.}  \nonumber
\end{eqnarray}
Note that in general the inequality
\begin{equation}
V_{1}V_{2}\geq \frac{a^{2}}{4}
\end{equation}
implies that
\begin{equation}
V_{1}+V_{2}\geq V_{1}+\frac{a^{2}}{4V_{1}}\geq a\text{,}
\end{equation}
so that the criterion of Eq.~(\ref{GGHI}) for {\it classical} teleportation
leads to
\begin{equation}
\Delta x_{\mu }^{2}+\Delta p_{\mu }^{2}\geq \frac{1}{2}\text{.}
\label{GGsum}
\end{equation}
Hence, the requirement of Grangier and Grosshans for{\it \ quantum}
teleportation is that
\begin{equation}
\Delta x_{\mu }^{2}+\Delta p_{\mu }^{2}<\frac{1}{2}\text{,}  \label{GGqtmu}
\end{equation}
which for $r\gg 1$ becomes
\begin{equation}
\Delta x^{2}+\Delta p^{2}<\frac{1}{2}\text{.}  \label{GGqt}
\end{equation}
Here, $(\Delta x^{2},\Delta p^{2})$ are as defined in Eq. (\ref{dxdpmu}),
now with $\mu =1$;
\begin{eqnarray}
\Delta x^{2} &=&\langle (x_{1}-x_{2})^{2}\rangle ,  \label{dxdp} \\
\Delta p^{2} &=&\langle (p_{1}+p_{2})^{2}\rangle \text{,}  \nonumber
\end{eqnarray}
where from Eq.~(\ref{wout}), we have that $\Delta x^{2}+\Delta p^{2}=\bar{%
\sigma}_{-}^{2}$ for the EPR\ beams $(1,2)$. The claim of Grangier and
Grosshans\cite{Grangier00a} is that the inequality of Eq.~(\ref{GGHI})
serves as ``the condition for no useful entanglement between the two
beams,'' where by `useful' they refer explicitly to ``the existence of
quantum non-separability [violation of Bell's inequalities].'' The
inequalities of Eqs.~(\ref{newHI}) and (\ref{GGHI}) are also related to
criteria developed within the setting of quantum nondemolition detection
(QND)\cite{Grangier98}, as discussed in the next section.

In a second paper\cite{Grangier00b}, Grangier and Grosshans introduce an
alternative criteria for the successful teleportation of coherent states,
namely that

\begin{quote}
``the information content of the teleported quantum state is higher than the
information content of any (classical or quantum) copy of the input state,
that may be broadcasted classically.''
\end{quote}

To quantify the concept of ``information content'' they introduce a
``generalized fidelity'' describing not the overlap of quantum states as is
standard in the quantum information community, but rather the conditional
probability $P(\alpha |I)$ that a particular coherent state $|\alpha \rangle
$ was actually sent given ``the available information $I$.'' In effect,
Grangier and Grosshans consider the following protocol. Victor sends to
Alice some unknown coherent state $|\alpha _{0}\rangle $, with Alice making
her best attempt to determine this state\cite{Arthurs65}, and sending the
resulting measurement outcome to Bob as in the standard protocol. Bob then
does one of two things. In the first instance, he forwards only this
classical message with Alice's measurement outcome to Victor without
reconstructing a quantum state. In the second case, he actually generates a
quantum state conditioned upon Alice's message and sends this state to
Victor, who must then make his own measurement to deduce whether the
teleported state corresponds to the one that he initially sent. For
successful teleportation, Grangier and Grosshans demand that the information
gained by Victor should be greater in the latter case where quantum states
are actually generated by Bob than in the former case where only Alice's
classical measurement outcome is distributed. It is straightforward to show
that Eq.~(\ref{GGqt}) given above is sufficient to ensure that this second
criteria is likewise satisfied for the teleportation of a coherent state $%
|\alpha \rangle $, albeit with the same caveat expressed in \cite{explainf},
namely that neither the set ${\it S}$ of initial states $\{|\psi
_{in}\rangle \}$ nor the distribution $P(|\psi _{in}\rangle )$ over these
states is specified.

We now turn to an evaluation of these criteria of Grangier and Grosshans
placing special emphasis on the issues of entanglement and violations of
Bell's inequalities, specifically because these are the concepts Grangier
and Grosshans emphasize in their work\cite{Grangier00a,Grangier00b}.

\section{Entanglement and Fidelity}

\subsection{Nonseparability of the EPR\ beams}

To address the question of the nonseparability of the EPR\ beams, we refer
to the papers of Duan {\it et al}.\ and of Simon\cite{duan00,simon00}, as
well as related work by Tan\cite{tan99}. For the definitions of $%
(x_{i},p_{i})$ that we have chosen for the EPR\ beams $(1,2)$, a sufficient
condition for nonseparability (without an assumption of Gaussian statistics)
is that
\begin{equation}
\Delta x^{2}+\Delta p^{2}<1\text{,}  \label{duancondition}
\end{equation}
where $\Delta x^{2}$ and $\Delta p^{2}$ are defined in Eq.~(\ref{dxdp}).
This result follows from Eq. (3) of Duan {\it et al.} with $a=1$ (and from a
similar more general equation in Simon)\cite{relate}. Note that Duan {\it et
al}.\ have $\Delta x_{i}^{2}=\frac{1}{2}=\Delta p_{i}^{2}$ for the vacuum
state, while our definitions lead to $\Delta x_{i}^{2}=\frac{1}{4}=\Delta
p_{i}^{2}$ for the vacuum state, where for example, $\Delta
x_{1}^{2}=\langle x_{1}{}^{2}\rangle $, and that all fields considered have
zero mean.

Given the Wigner distribution $W_{{\rm EPR}}^{out}$ as in Eq.~(\ref{wout}),
we find immediately that
\begin{eqnarray}
\Delta x^{2}+\Delta p^{2} &=&2\frac{\bar{\sigma}_{-}^{2}}{2}
\label{separable} \\
&=&\eta e^{-2r}+(1-\eta )(1+2\bar{n})\text{. }  \nonumber
\end{eqnarray}
For the case $\bar{n}=0$, the resulting state is {\it always entangled for
any }$r>0$ {\it even for }$\eta \ll 1$, in agreement with the discussion in
Duan {\it et al.}\cite{duan00}. For nonzero $\bar{n}$, the state is
entangled so long as
\begin{equation}
\bar{n}<\frac{\eta \lbrack 1-\exp (-2r)]}{2(1-\eta )}\text{.}  \label{nlimit}
\end{equation}

We emphasize that in the experiment of Furusawa {\it et al.}\cite{Furusawa98}
for which $\bar{n}=0$ is the relevant case, the above inequality guarantees
that teleportation was carried out with entangled (i.e., nonseparable)
states for the EPR beams, independent of any assumption about whether these
beams were Gaussian or pure states\cite{expvalues}.

By contrast to the condition for entanglement given in Eq.~(\ref
{duancondition}), Grangier and Grosshans require instead the more stringent
condition of Eq.~(\ref{GGqtmu}) for successful teleportation. Although they
would admit that the EPR beams are indeed entangled whenever Eq.~(\ref
{duancondition}) is satisfied,\footnote{%
Grangier was in fact unaware of Refs.\cite{duan00,simon00} when Ref.\cite
{Grangier00a} was originally posted, having had this work pointed out by us.}
they would term entanglement in the domain
\[
\frac{1}{2}\leq \Delta x^{2}+\Delta p^{2}<1\text{ }
\]
as not ``useful''\cite{Grangier00a}.

With regard to the QND-like conditions introduced by Grangier and Grosshans
\cite{Grangier00a}, we note that more general forms for the nonseparability
condition of Eq.~(\ref{duancondition}) are given in Refs.\cite
{duan00,simon00}. Of particular relevance is a condition for the variances
of Eq.~(\ref{munuvar}) for the case of symmetric fluctuations as for EPR
state in Eq.~(\ref{wout}), $\mu _{ij}=\mu _{ji}\equiv \mu $ and $\nu
_{ij}=\nu _{ji}\equiv \nu $, with $\mu =-\nu $. Consider for example the
first set of variances in Eq. (\ref{dxdpmu}), namely
\begin{equation}
\Delta x_{\mu }^{2}=\langle (x_{2}-\mu x_{1})^{2}\rangle \text{ \ \ \ and \
\ \ }\Delta p_{\mu }^{2}=\langle (p_{2}+\mu p_{1})\rangle ^{2}\text{,}
\label{xpinfer}
\end{equation}
as would be appropriate for an inference of $(x_{2},p_{2})$ from a
measurement (at a distance) of $(x_{1},p_{1})$. In this case, a sufficient
condition for entanglement of the EPR beams $(1,2)$ 
may be obtained using Eq.~(11) of Ref.~\onlinecite{simon00} yielding
\begin{equation}
\Delta x_{\mu }^{2}+\Delta p_{\mu }^{2}<\frac{(1+\mu ^{2})}{2}\text{,}
\label{muduancondition}
\end{equation}
which reproduces Eq.~(\ref{duancondition}) for $\mu =1$. This equation for
nonseparability implies that
\begin{equation}
\Delta x_{\mu }^{2}\Delta p_{\mu }^{2}<\frac{(1+\mu ^{2})^{2}}{16}\text{,}
\label{HImu}
\end{equation}
which is in the form of a Heisenberg-type inequality. Note that this
inequality is satisfied for any $r>0$ and $0<\eta \leq 1$ for $\bar{n}=0$.
As discussed in Refs.\cite{Ou92,reid89}, $\mu $ must be chosen in
correspondence to the degree of correlation between the EPR beams, with $%
0<\mu \leq 1$. An explicit expression for our current model given in Eq. (%
\ref{muopt}). By constrast, in applying their QND-like conditions, Grangier
and Grosshans demand to the contrary the Heisenberg-type inequality
\begin{equation}
\Delta x_{\mu }^{2}\Delta p_{\mu }^{2}<\frac{1}{16}\text{.}  \label{HIGGmu}
\end{equation}
Within the setting our current model, this condition can only be satisfied
for efficiency $\eta >\frac{1}{2}$ \cite{PGprivate}. 
Although this criterion has been found to
be useful in the analysis of back-action evading measurement for quantum
nondemolition detection, it apparently has no direct relevance to the
question of entanglement, for $\mu =1$ or otherwise.

Certainly, $\mu =1$ is the case relevant to the actual teleportation protocol
of Ref.\cite{Braunstein98c}. However, Alice and Bob are surely free to
explore the degree of correlation between their EPR beams and to test for
entanglement by any means at their disposal, including simple measurements
with $\mu \neq 1$.

Although the boundary expressed by the nonseparability conditions of Eqs.~(%
\ref{duancondition}) and (\ref{muduancondition}) are perhaps not so familiar
in Quantum Optics, we stress that these criteria are associated quite
directly with the standard condition for nonclassical behavior adopted by
this community. Whenever Eqs.~(\ref{duancondition}) and (\ref
{muduancondition}) are satisfied, the Glauber-Sudarshan phase-space function
becomes non-positive\cite{simon00-gs}, which for almost forty years has
heralded entrance into a manifestly quantum or nonclassical domain. It is
difficult to understand how Grangier and Grosshans propose to move from $%
\Delta x^{2}+\Delta p^{2}=1$ to $\Delta x^{2}+\Delta p^{2}=\frac{1}{2}$
without employing quantum resources in the teleportation protocol (as is
required when the Glauber-Sudarshan $P$-function is not positive definite).
Their own work offers no suggestion of how this is to be accomplished.

\subsection{Fidelity}

Turning next to the question of the relationship of entanglement of the EPR
beams [as quantified in Eq.~(\ref{duancondition})] to the fidelity
attainable for teleportation {\it with these beams}, we recall from Eq. (2)
of Ref.\cite{Furusawa98} that
\begin{equation}
F=\frac{1}{1+\bar{\sigma}_{-}^{2}}\text{,}  \label{fscience}
\end{equation}
where this result applies to teleportation of coherent states 
\cite{notation,alphabet}.
When combined with Eq.~(\ref{separable}), we find that
\begin{equation}
F=\frac{1}{1+(\Delta x^{2}+\Delta p^{2})}\text{,}  \label{fidelityduan}
\end{equation}
The criterion of Eq.~(\ref{duancondition}) for nonseparability then
guarantees that nonseparable EPR states as in Eqs.~(\ref{wepr},\ref{wout}%
)(be they mixed or pure) are sufficient to achieve
\begin{equation}
F>F_{classical}=\frac{1}{2}\text{,}  \label{fidelitylimit}
\end{equation}
whereas separable states must have $F\leq F_{classical}=\frac{1}{2}$,
although we emphasize that this bound applies for the average fidelity for
coherent states distributed over the entire complex plane\cite
{Fuchs00,alphabet}.

{\it We thereby demonstrate that the condition }$F>F_{classical}=\frac{1}{2}$
{\it \ for quantum teleportation as established in Ref.\cite{Fuchs00}
coincides with that for nonseparability (i.e., entanglement) of Refs.\cite
{duan00,simon00} for the EPR state of Eq.~(\ref{wout}). }Note that for $\bar{%
n }=0$, we have
\begin{equation}
F=\frac{1}{2-\eta (1-e^{-2r})}\text{,}  \label{freta}
\end{equation}
so that the entangled EPR beams considered here (as well as in Refs.\cite
{Grangier00a,Grangier00b}) provide a sufficient resource for beating the
limit set by a classical channel alone for any $r>0$, so long as $\eta >0$.
In fact, the quantities $(\Delta x^{2},\Delta p^{2})$ are readily measured
experimentally, so that the entanglement of the EPR\ beams can be
operationally verified, as was first accomplished in Ref.\cite{Ou92}, and
subsequently in Ref.\cite{Furusawa98}. We stress that independently of any
further assumption, the condition of Eq.~(\ref{duancondition}) is sufficient
to ensure entanglement for pure or mixed states\cite{necandsuff,vanLoock00b}.

The dependence of fidelity $F$ on the degree of squeezing $r$ and efficiency
$\eta $ as expressed in Eq.~(\ref{freta}) is illustrated in Figure 1. Here,
in correspondence to an experiment with fixed overall losses and variable
parametric gain in the generation of the EPR entangled state, we show a
family of curves in Figure 1 each of which is drawn for constant $\eta $ as
a function of $r$. Clearly, $F>F_{classical}=\frac{1}{2}$ and hence
nonseparability results in each case. The only apparent significance of $F=%
\frac{2}{3}$ as championed by Grangier and Grosshans (and which results for $%
\Delta x^{2}+\Delta p^{2}=\frac{1}{2}$) is to bound $F$ for $\eta =0.5$.

\begin{figure}[h]
 \centering
  \epsfxsize=8cm \epsfbox{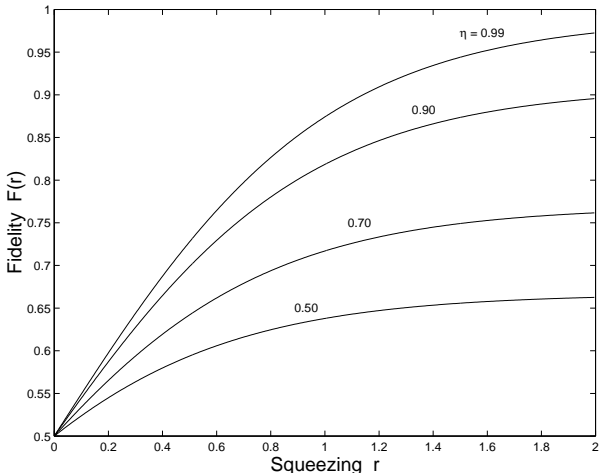}
\caption{\protect\narrowtext Fidelity $F$ as given by Eq.~(\protect\ref{freta})
versus the degree of squeezing $r$ for fixed efficiency $\protect\eta $. 
From top to bottom, the curves are drawn with 
$\protect\eta =\protect\{0.99,0.90,0.70,0.50\protect\}$ in
correspondence to increasing loss $(1-\protect\eta )$. Note that 
$F_{classical}=\protect\frac{1}{2}$ provides a demarcation between 
separable and nonseparable states (mixed or otherwise), while 
$F=\protect\frac{2}{3}$ is apparently of no particular significance, the 
contrary claims of Ref.~[11,12] notwithstanding. Note that for 
$\protect\eta =1$, $r=\protect\frac{\protect\ln 2}{2}=0.3466$ gives 
$F=\protect\frac{2}{3}$, corresponding to $-3$dB of squeezing. In all
cases, $\protect\bar{n}=0$.}
\label{figure1}
\end{figure}

As for the criterion of ``information content'' introduced by Grangier and
Grosshans\cite{Grangier00b}, we note it can be easily understood from the
current analysis and the original discussion in Ref.\cite{Braunstein98c}.
Each of the interventions by Alice and Bob represent one unit of added
vacuum noise that will be convolved with the initial input state in the
teleportation protocol (the so-called {\it quduties}). Grangier and
Grosshans compare the following two situations: (i) Bob passes directly the
classical information that he receives to Victor and (ii) Bob generates a
quantum state in the usual fashion that is then passed to Victor. Grangier
and Grosshans would demand that Victor should receive the same information
in these two cases, which requires that $\bar{\sigma}_{-}^{2}=\Delta
x^{2}+\Delta p^{2}<\frac{1}{2}$, and hence $F>\frac{2}{3}$. That is, as the
degree of correlation between the EPR\ beams is increased, there comes a
point for which $\Delta x^{2}+\Delta p^{2}=\frac{1}{2}$, and for which each
of Alice and Bob's excess noise has been reduced from $1$ quduty each to $%
\frac{1}{2}$ quduty each. At this point, Grangier and Grosshans would
(arbitrarily) assign the entire resulting noise of $\frac{1}{2}+\frac{1}{2}%
=1 $ quduties to Alice, with then the perspective that Bob's state
recreation adds no noise. Of course one could equally well make the
complementary assignment, namely $1$ quduty to Bob and none to Alice (again
in the case with $\bar{\sigma}_{-}^{2}=\frac{1}{2}$). The point that seems
to be missed by Grangier and Grosshans is that key to quantum teleportation
is the transport of quantum states. Although they correctly state that
``there is {\it no} extra noise associated to the reconstruction: given a
measured $\beta $, one can exactly reconstruct the coherent state $|\beta
\rangle $, by using a deterministic translation of the vacuum.'' Bob can
certainly make such a state deterministically, but it is an altogether
different matter for Victor to receive a classical number from Bob in case
(i) as opposed to the actual quantum state in (ii). In this latter case
apart from having a physical state instead of a number, Victor must actually
make his own measurement with the attendant uncertainties inherent in $%
|\beta \rangle $ then entering. Analogously, transferring measurement
results about a qubit, without recreating a state at the output (i.e.,
without sending an actual {\it quantum state\/} to Victor), is not what is
normally considered to constitute quantum teleportation relative to the
original protocol of Bennett {\it et al.}\cite{Bennett93}.

Turning next to the actual experiment of Ref.\cite{Furusawa98}, we note that
a somewhat subtle issue is that the detection efficiency for Alice of the
unknown state was not $100$\%, but rather was $\eta _{A}^{2}=0.97$. Because
of this, the fidelity for classical teleportation (i.e., with vacuum states
in place of the EPR beams) did not actually reach $\frac{1}{2}$, but was
instead $F_{0}=0.48$. This should not be a surprise, since there is nothing
to ensure that a given classical scheme will be optimal and actually reach
the bound $F_{classical}=\frac{1}{2}$. Hence, the starting point in the
experiment with $r=0$ had $F_{0}<F_{classical}$; the EPR beams with $r>0$
(which were in any event entangled by the above inequality) then led to
increases in fidelity from $F_{0}$ upward, exceeding the classical bound $%
F_{classical}=\frac{1}{2}$ for a small (but not infinitesimal) degree of
squeezing. Note that the whole effect of the offset $F_{0}=0.48<\frac{1}{2}$
can be attributed to the lack of perfect (homodyne) efficiency at Alice's
detector for the unknown state. In the current discussion for determining
the classical bound in the {\it optimal} case, we set Alice's detection
efficiency instead to $\eta _{A}^{2}=1$, then as shown above, classical
teleportation will achieve $F=\frac{1}{2}$.

Independent of such considerations, we reiterate that the nonseparability
condition of Refs.\cite{duan00,simon00} applied to the EPR state of Eqs.~(%
\ref{wepr}) and (\ref{wout}) leads to the same result $F_{classical}=\frac{1%
}{2}$ [Eqs.~(\ref{fidelityduan}) and (\ref{fidelitylimit})] as did our
previous analysis based upon teleportation with only a classical
communication channel linking Alice and Bob\cite{Fuchs00}. This convergence
further supports $F_{classical}=\frac{1}{2}$ as the appropriate
quantum-classical boundary for the teleportation of coherent states, the
claims of Grangier and Grosshans notwithstanding. Relative to the original
work of Bennett {\it et al.}\cite{Bennett93}, exceeding the bound $%
F_{classical}=\frac{1}{2}$ for the teleportation of coherent can be
accomplished with a classical channel and entangled (i.e., nonseparable) EPR
states, be they mixed or pure, as is made clear by the above analysis and as
has been operationally confirmed\cite{Furusawa98}.

We should however emphasize that the above conclusions concerning
nonseparability and teleportation fidelity apply to the specific case of the
EPR state as in Eq. (\ref{wout}), for which inequality Eq. (\ref
{duancondition}) represents both a necessary and sufficient criterion for
nonseparability according to Refs.\cite{duan00,simon00}. More generally, for
arbitrary entangled states, nonseparability does not necessarily lead to $F>%
\frac{1}{2}$ in coherent-state teleportation\cite{necandsuff,vanLoock00b}.

\section{Bell's Inequalities}

The papers by Banaszek and Wodkiewicz\cite{b-k98,b-k99} provide our point of
reference for a discussion of Bell's inequalities. In these papers, the
authors introduce an appropriate set of measurements that lead to a Bell
inequality of the CHSH type. More explicitly, Eq.(4) of Ref.\cite{b-k98}
gives the operator $\hat{\Pi}(\alpha ;\beta )$ whose expectation values are
to be measured. Banaszek and Wodkiewicz point out that the expectation value
of $\hat{\Pi}(\alpha ;\beta )$ is closely related the Wigner function of the
field being investigated, namely
\begin{equation}
W(\alpha ;\beta )=\frac{4}{\pi ^{2}}\Pi (\alpha ;\beta )\text{ ,}
\label{bigpi}
\end{equation}
where $\Pi (\alpha ;\beta )=\langle \hat{\Pi}(\alpha ;\beta )\rangle $.

For the entangled state shared by Alice and Bob in the teleportation
protocol, we identify $W_{{\rm EPR}}^{out}$ as the relevant Wigner
distribution for the modes $(1,2)$ of interest, so that
\begin{eqnarray}
&&\Pi _{{\rm EPR}}^{out}(x_{1},p_{1};x_{2},p_{2})  \label{piepr} \\
&=&\frac{1}{\bar{\sigma}_{+}^{2}\bar{\sigma}_{-}^{2}}\exp
\{-[(x_{1}+x_{2})^{2}+(p_{1}-p_{2})^{2}]/\bar{\sigma}_{+}^{2}  \nonumber \\
&&~-[(x_{1}-x_{2})^{2}+(p_{1}+p_{2})^{2}]/\text{ }\bar{\sigma}_{-}^{2}\}
\text{ \ .}  \nonumber
\end{eqnarray}
Banaszek and Wodkiewicz show that $\Pi _{{\rm EPR}
}^{out}(x_{1},p_{1};x_{2},p_{2})$ gives directly the correlation function
that would otherwise be obtained from a particular set of observations over
an ensemble representing the field with density operator $\hat{\rho}$, where
the actual measurements to be made are as described in Refs.\cite
{b-k98,b-k99}. In simple terms, $\hat{\Pi}_{{\rm EPR}}^{out}(0,0;0,0)$ is
the parity operator for separate measurements of photon number on modes $%
(1,2)$, with then nonzero $(x_{i},p_{i})$ corresponding to a ``rotation'' on
the individual mode $i$ that precedes its parity measurement.

\begin{figure}[h]
\centering
 \caption{\protect\narrowtext The function ${\protect\cal B}({\protect\cal J})$
from Eq.~(\protect\ref{bdefined}) as a function of ${\protect\cal J}$ for 
various values of $(r,\protect\eta )$. Recall that ${\protect\cal B}>2$ 
heralds a direct violation of the CHSH inequality, with the dashed 
line ${\protect\cal B}=2$ shown. In each of the plots (a)-(d) a family of 
curves is drawn for fixed efficiency $\protect\eta $ and
four values of 
$r=\protect\{0.1,\protect\frac{\protect\ln 2}{2},1.0,2.0\protect\}$. 
(a) $\protect\eta =0.99$, (b) $\protect\eta =0.90$, (c) $\protect\eta =0.70$, 
(d) $\protect\eta =0.50 $; in all cases, $\protect\bar{n}=0$.} 
\centering
  \epsfxsize=7.2cm \epsfbox{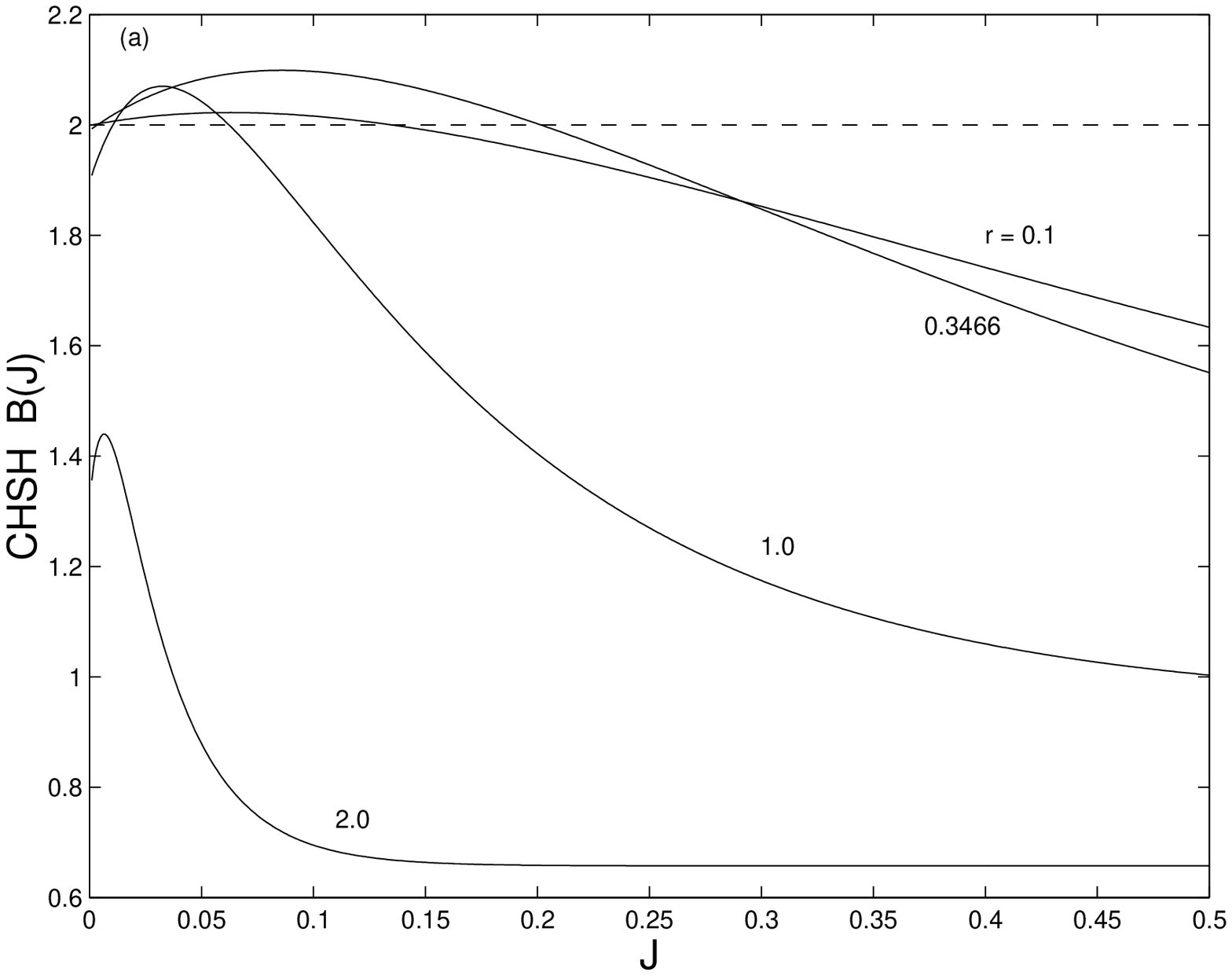}
  \epsfxsize=7.2cm \epsfbox{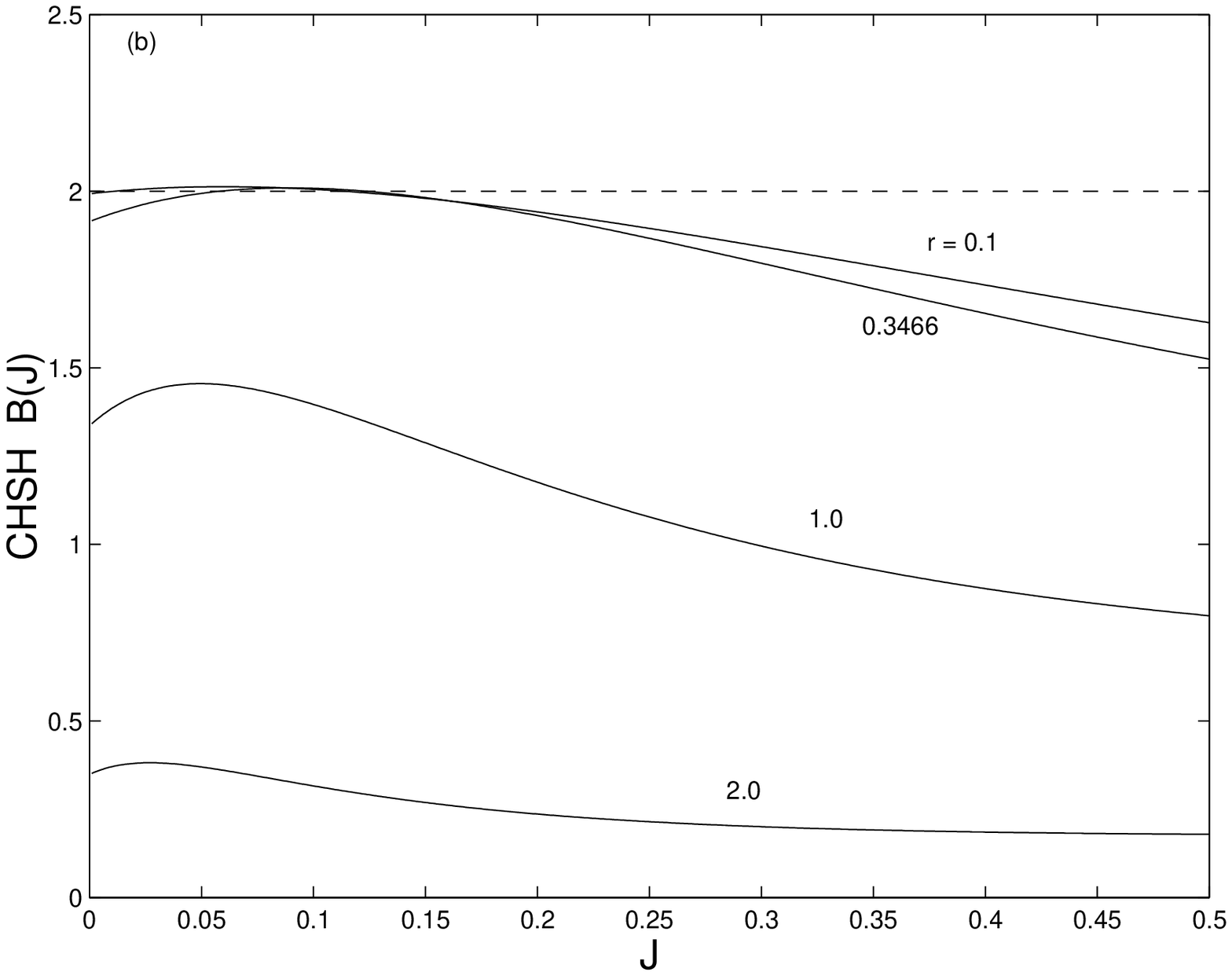}
  \epsfxsize=7.2cm \epsfbox{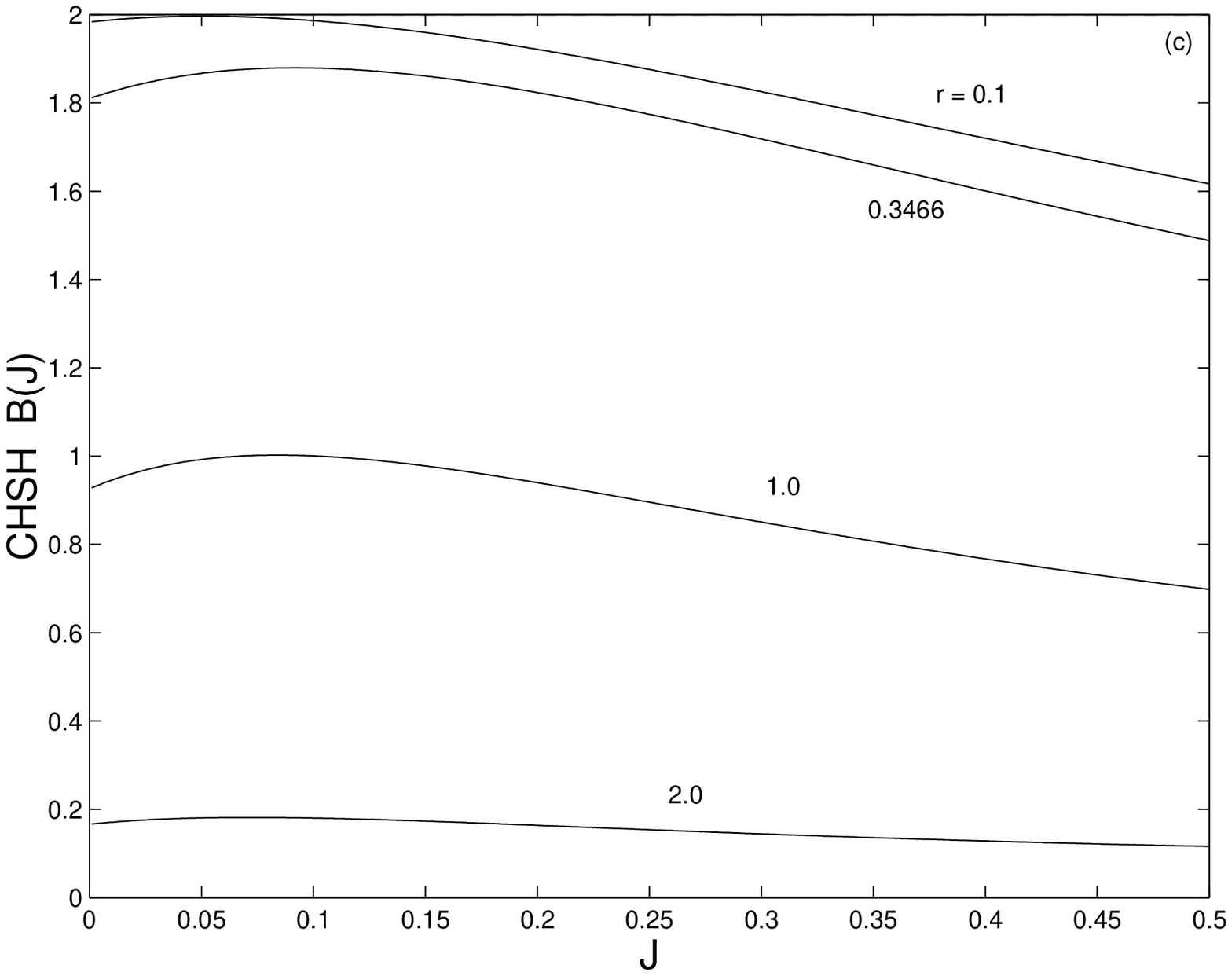}
  \epsfxsize=7.2cm \epsfbox{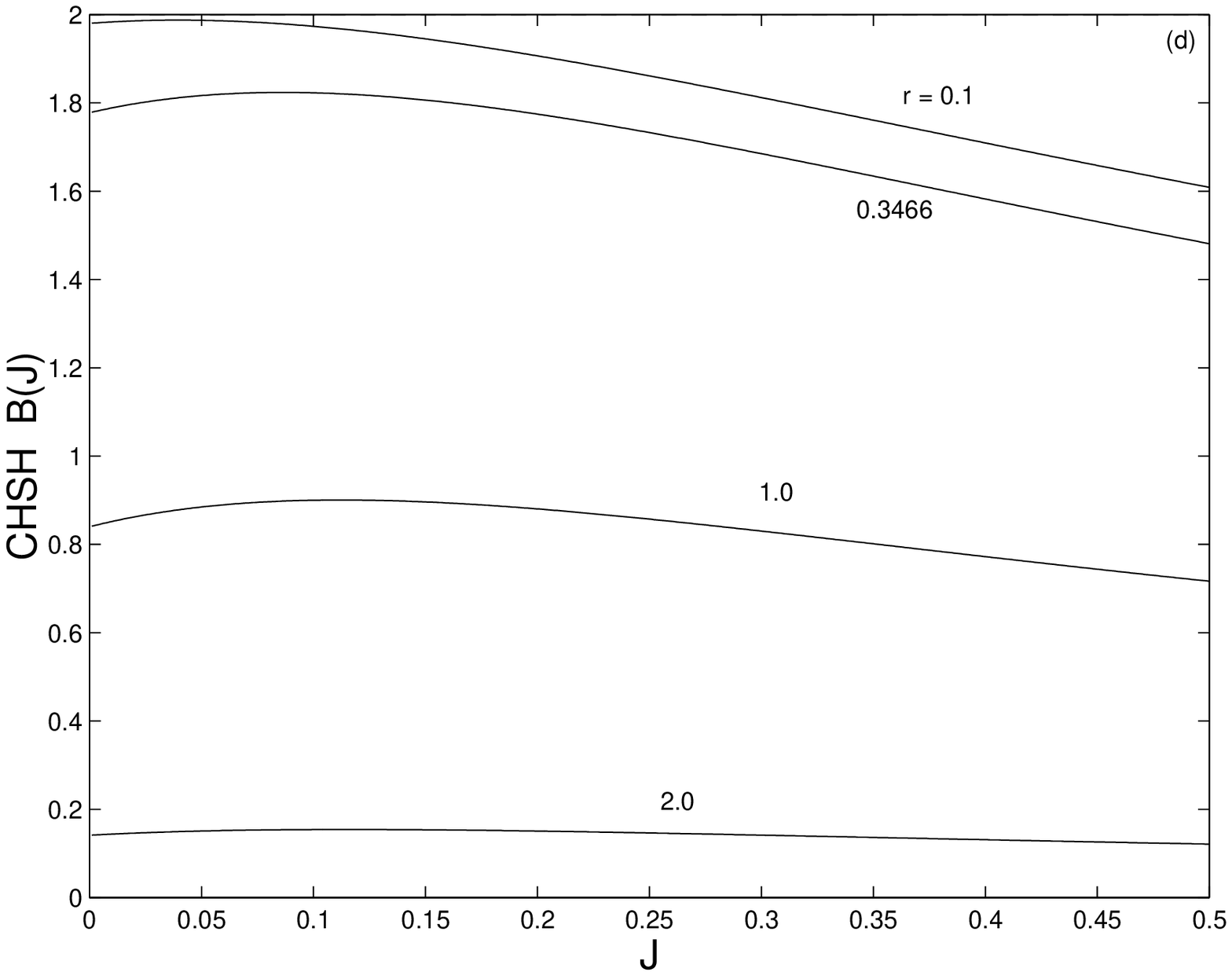}
\label{figure2}
\end{figure}

The function constructed by Banaszek and Wodkiewicz to test for local hidden
variable theories is denoted by ${\cal B}$ and is defined by
\begin{eqnarray}
&&{\cal B}({\cal J})  \label{bdefined} \\
&=&\Pi _{{\rm EPR}}^{out}(0,0;0,0)+\Pi _{{\rm EPR}}^{out}(\sqrt{{\cal J}}
,0;0,0)  \nonumber \\
&&+\Pi _{{\rm EPR}}^{out}(0,0;-\sqrt{{\cal J}},0)-\Pi _{{\rm EPR}}^{out}(
\sqrt{{\cal J}},0;-\sqrt{{\cal J}},0)\text{ ,}  \nonumber
\end{eqnarray}
where ${\cal J}$ is a positive (real) constant. As shown in Ref.\cite
{b-k98,b-k99}, any local theory must satisfy
\begin{equation}
-2\leq {\cal B}\leq 2\text{ .}  \label{bi}
\end{equation}
As emphasized by Banaszek and Wodkiewicz for the lossless case, $\Pi _{{\rm %
EPR}}^{out}(0,0;0,0)=1$ ``describes perfect correlations ... as a
manifestation of ... photons always generated in pairs.''

There are several important points to be made about this result. In the
first place, in the ideal case with no loss $(\eta =1)$, there is a
violation of the Bell inequality of Eq.~(\ref{bi}) for any $r>0$. Further,
this threshold for the onset of violations of the CHSH inequality coincides
with the threshold for entanglement as given in Eq.~(\ref{duancondition}),
which likewise is the point for surpassing $F_{classical}=\frac{1}{2}$ as in
Eqs. (\ref{fidelityduan},\ref{fidelitylimit}) and as shown in our prior
analysis of Ref.\cite{Fuchs00} which is notably based upon a quite different
approach.

Significantly, there is absolutely nothing special about the point $r=\frac{%
\ln 2}{2}\approx 0.3466$ (i.e., the point for which $\exp [-2r]=0.5$ and for
which $F=\frac{2}{3}$ for the teleportation of coherent states), in
contradistinction to the claims of Grangier and Grosshans to the contrary
\cite{Grangier00a,Grangier00b}. Instead, any $r>0$ leads to a nonseparable
EPR state, to a violation of a Bell inequality, and to $F>F_{classical}=%
\frac{1}{2}$ for the teleportation of coherent states. There is certainly no
surprise here since we are dealing with pure states for $\eta =1$\cite
{Peres92}.

We next examine the case with $\eta <1$, which is clearly of interest for
any experiment. Figure 2 illustrates the behavior of ${\cal B}$ as a
function of ${\cal J}$ for various values of the squeezing parameter $r$ and
of the efficiency $\eta $. Note that throughout our analysis in this
section, we make no attempt to search for optimal violations, but instead
follow dutifully the protocol of Banaszek and Wodkiewicz as expressed in
Eq.~(\ref{bdefined}) for the case with losses as well.

From Figure 2 we see that for any particular set of parameters $(r,\eta )$,
there is an optimum value ${\cal J}_{\max }$ that leads to a maximum value
for ${\cal B}({\cal J}_{\max })$, which is a situation analogous to that
found in the discrete variable case. By determining the corresponding value $%
{\cal J}_{\max }$ at each $(r,\eta )$, in Figure 3 we construct a plot that
displays the dependence of ${\cal B}$ on the squeezing parameter $r$ for
various values of efficiency $\eta $. Note that all cases shown in the
figure lead to fidelity $F>F_{classical}$.

\begin{figure}[h]
\centering
 \epsfxsize=8cm \epsfbox{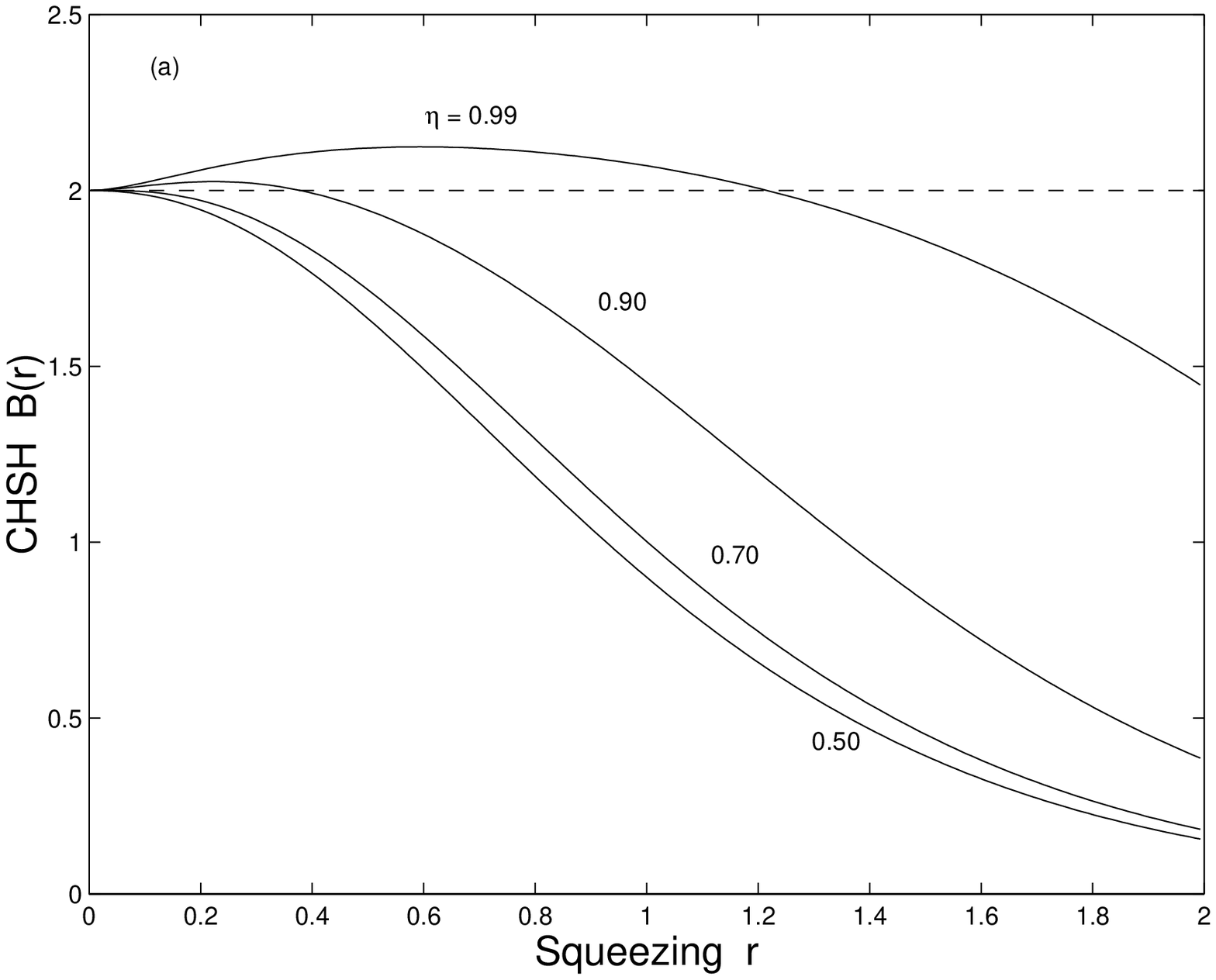}
 \epsfxsize=8cm \epsfbox{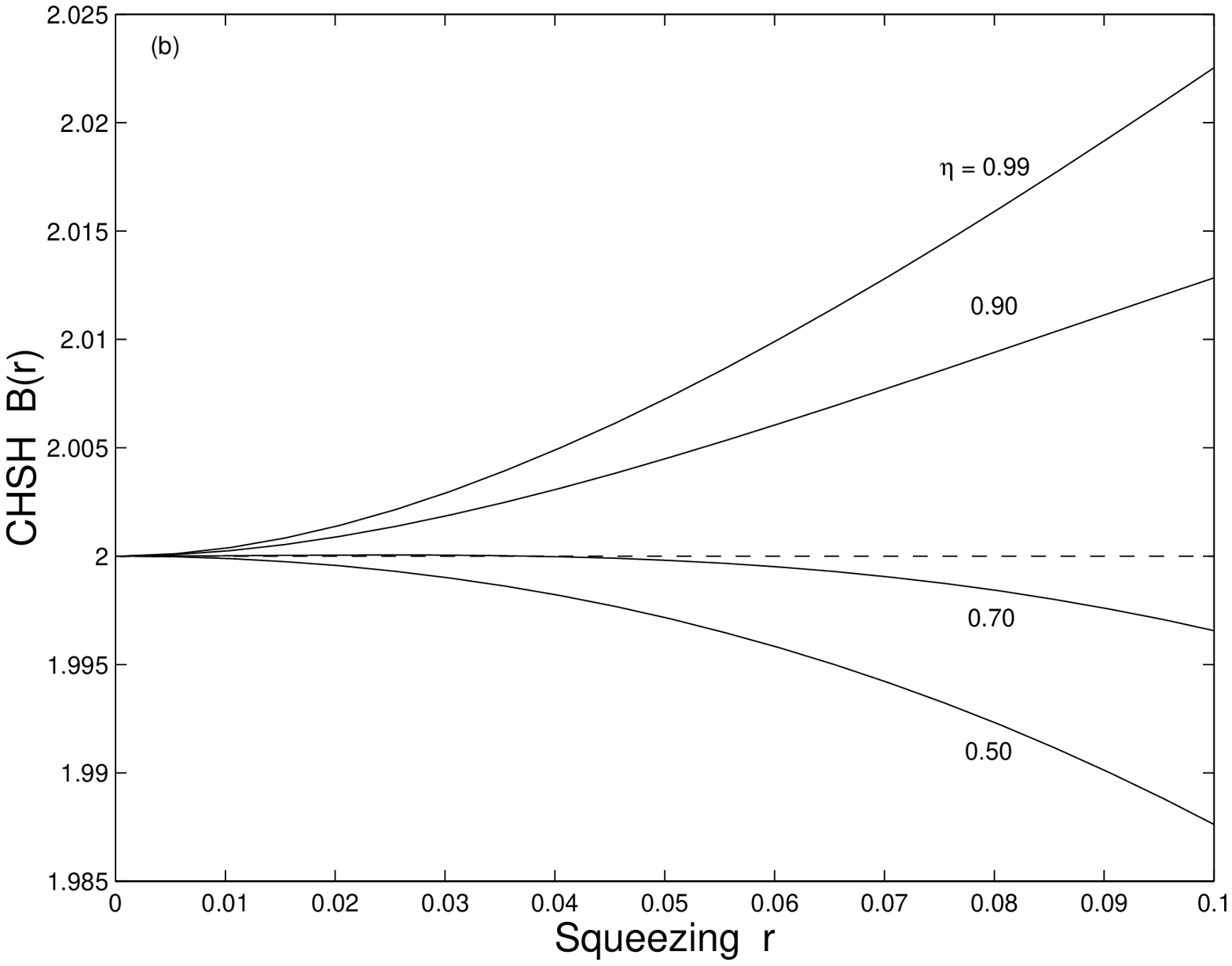}
\caption{\protect\narrowtext (a) The quantity ${\protect\cal B}$ from 
Eq.~(\protect\ref{bdefined}) as a function of $r$ for various values of 
efficiency $\protect\eta =\protect\{0.99,0.90,0.70,0.50\protect\}$ as 
indicated. At each point in $(r,\protect\eta )$, the value of 
${\protect\cal J}$ that maximizes ${\protect\cal B}$ has been chosen. Recall
that ${\protect\cal B}>2$ heralds a direct violation of the CHSH inequality, 
with the dashed line ${\protect\cal B}=2$ shown. Also note that 
$F>\protect\frac{1}{2}$ for all $r>0$. (b) An expanded view of 
${\protect\cal B}$ in the small $r$ region $r\protect\leq 0.1$. Note that in 
the case $\protect\eta =0.70$, ${\protect\cal B}>2 $ for small $r$.
In all cases, $\protect\bar{n}=0$.}
\label{figure3}
\end{figure}

For $\frac{2}{3}<\eta \leq 1$ there are regions in $r$ that produce direct
violations of the Bell inequality considered here, namely ${\cal B}>2$\cite
{etatwothirds}. In general, these domains with ${\cal B}>2$ contract toward
smaller $r$ with increasing loss $(1-\eta )$. In fact as $r$ increases, $%
\eta $ must become very close to unity in order to preserve the condition $%
{\cal B}>2$, where for $r\gg 1$,
\begin{equation}
2(1-\eta )\cosh (2r)\ll 1\text{.}  \label{bcondition}
\end{equation}
This requirement is presumably associated with the EPR state becoming more
``nonclassical'' with increasing $r$ and hence more sensitive to dissipation
\cite{Jeong00}. Stated somewhat more quantitatively, recall that the
original state $|EPR\rangle _{{\rm \scriptscriptstyle}1,2}$ of Eq.~(\ref
{eprstate}) is expressed as a sum over correlated photon numbers for each of
the two EPR beams $(1,2)$. The determination of ${\cal B}$ derives from
(displaced) parity measurements \ on the beams $(1,2)$ (i.e., projections
onto odd and even photon number), so that ${\cal B}$ should be sensitive to
the loss of a single photon. The mean photon number $\bar{n}_{i}$ for either
EPR beam goes as $\sinh ^{2}r$, with then the probability of losing no
photons after encountering the beam-splitter with transmission $\eta $
scaling as roughly $p_{0}\sim \lbrack \eta ]^{\bar{n}_{i}}$. We require that
the total probability for the loss of one or more photons to be small, so
that
\begin{equation}
(1-p_{0})\ll 1\text{,}
\end{equation}
and hence for $(1-\eta )\ll 1$ and $r\gg 1$ that
\begin{equation}
(1-\eta )\bar{n}_{i}\sim (1-\eta )\exp (2r)\ll 1\text{,}
\end{equation}
in correspondence to Eq.~(\ref{bcondition})\cite{alternate}.

On the other hand, note that small values of $r$ in Figure 3 lead to direct
violations of the CHSH inequality ${\cal B}>2$ with much more modest
efficiencies\cite{Jeong00}. In particular, note that for $r=\frac{\ln 2}{2}%
\approx 0.3466$ and $\eta =0.90$, $F<\frac{2}{3}$ [from Eq.~(\ref{freta})].
This case and others like it provide examples for which mixed states are
nonseparable and yet directly violate a Bell inequality, but for which $%
F\leq \frac{2}{3}$. Such mixed states do not satisfy the criteria of
Grangier and Grosshans (neither with respect to their Heisenberg-type
inequality nor with respect to their information exchange), yet they are
states for which $\frac{1}{2}<$ $F\leq \frac{2}{3}$ and ${\cal B}>2$, which
in and of itself calls the claims of Grangier and Grosshans into question.
There remains the possibility that $F>\frac{2}{3}$ might be sufficient to
warranty that mixed states in this domain would satisfy that ${\cal B}>2$,
and hence to exclude a description of the EPR\ state in terms of a local
hidden variables theory.

To demonstrate that this is emphatically not the case, we examine further
the relationship between the quantity ${\cal B}$ relevant to the CHSH
inequality and the fidelity $F$. Figure 4 shows a parametric plot of ${\cal B%
}$ versus $F$ for various values of the efficiency $\eta $. The curves in
this figure are obtained from plots as in Figures 1 and 3 by eliminating the
common dependence on $r$. From Figure 4, we are hard pressed to find any
indication that the value $F=\frac{2}{3}$ is in any fashion noteworthy with
respect to violations of the CHSH inequality. In particular, for efficiency $%
\eta \simeq 0.90$ most relevant to current experimental capabilities, the
domain $F>\frac{2}{3}$ is one largely devoid of instances with ${\cal B}>2$,
in contradistinction to the claim of Grangier and Grosshans that this domain
is somehow ``safer''\cite{Grangier00a} with respect to violations of Bell's
inequalities. Moreover, contrary to their dismissal of the domain $\frac{1}{2%
}<F\leq \frac{2}{3}$ as not being manifestly quantum, we see from Figure 4
that there are in fact regions with ${\cal B}>2$. Overall, the conclusions
of Grangier and Grosshans\cite{Grangier00a} related to the issues of
violation of a Bell inequality and of teleportation fidelity are simply not
supported by an actual quantitative analysis.

\begin{figure}[h]
\centering
 \epsfxsize=8cm \epsfbox{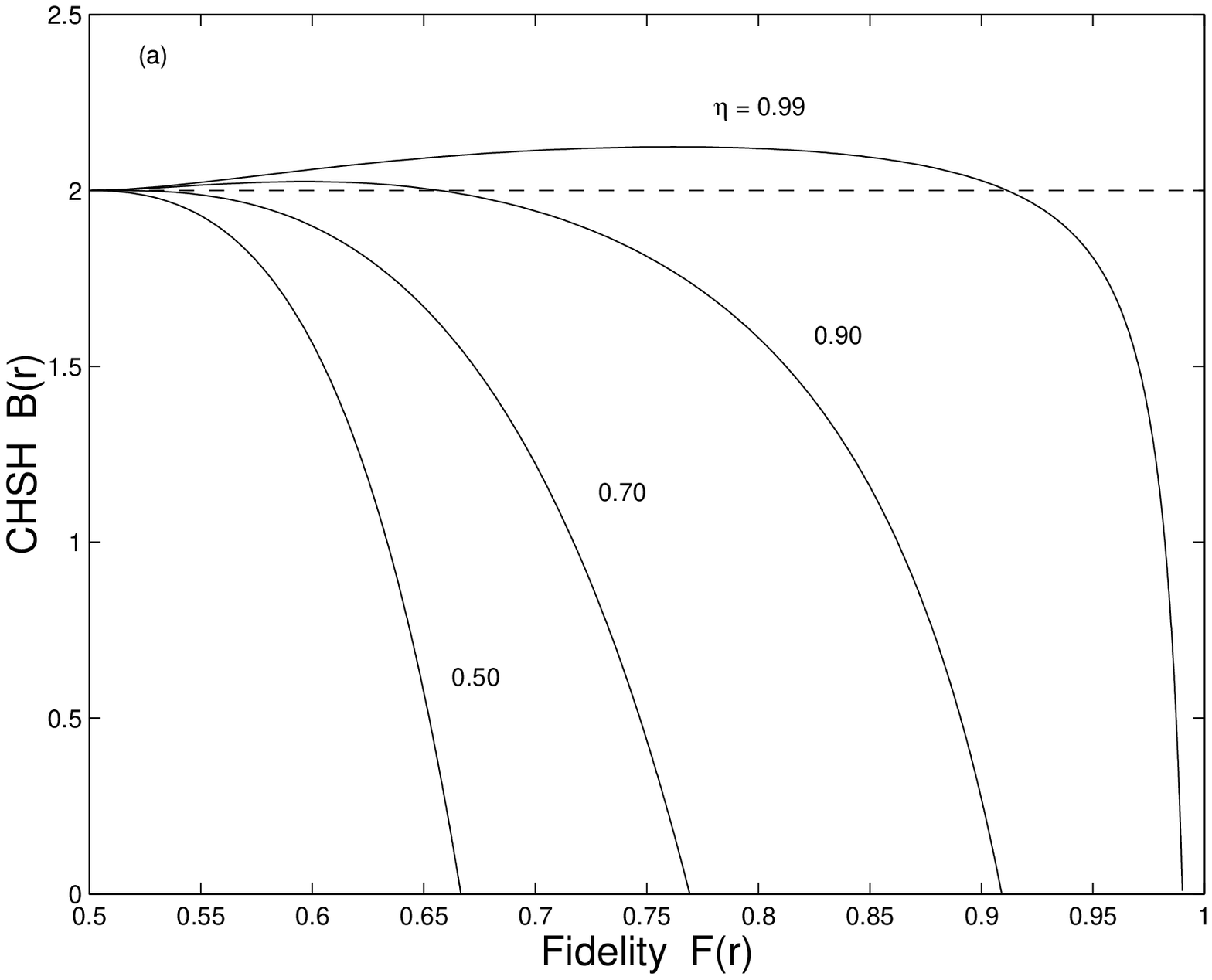}
 \epsfxsize=8cm \epsfbox{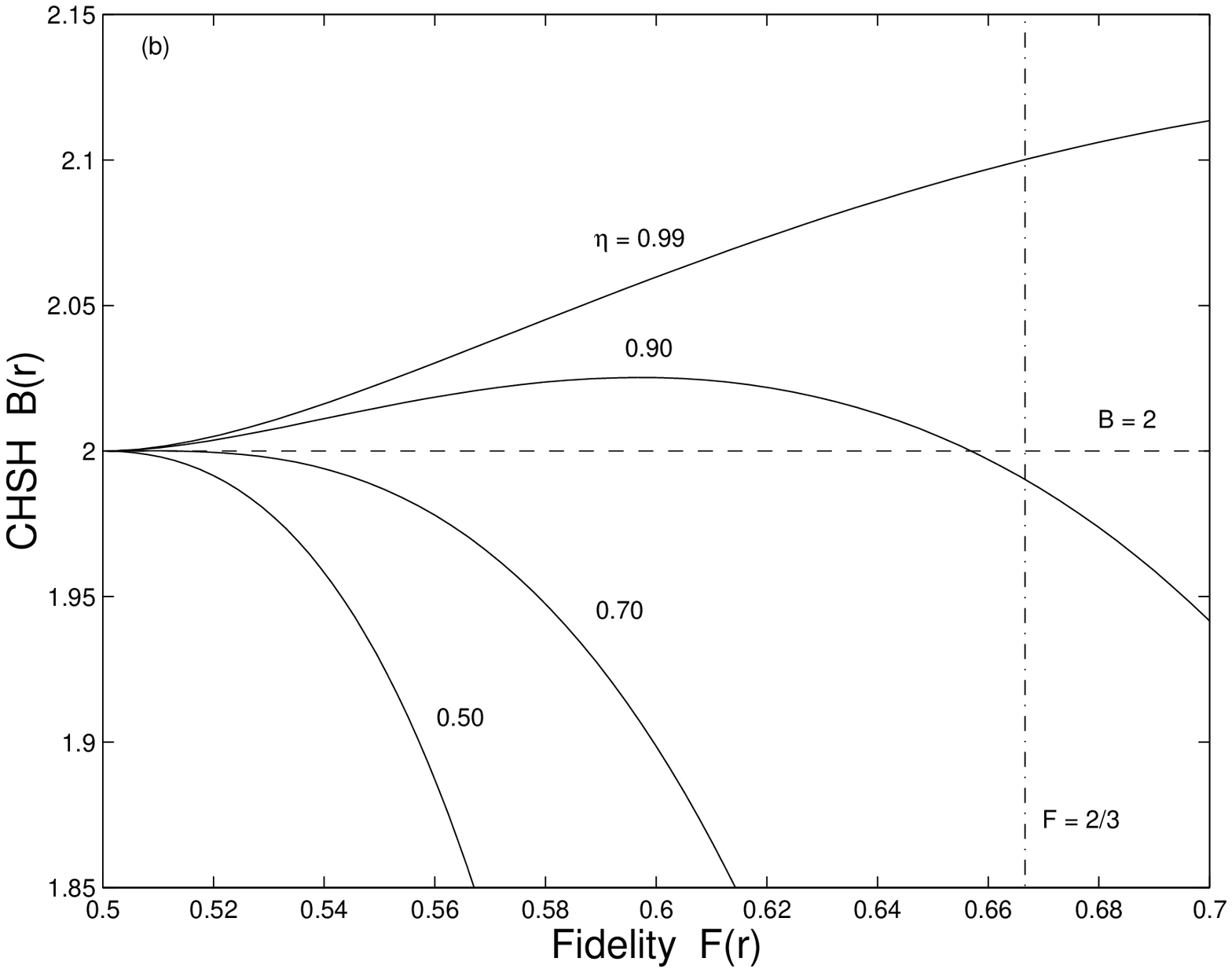}
\caption{\protect\narrowtext (a) A parametric plot of the CHSH quantity 
${\protect\cal B}$ [Eq.~(\protect\ref{bdefined})] versus fidelity $F$ 
[Eq.~(\protect\ref{freta})]. The curves are constructed from Figures 1 
and 3 by eliminating the $r$ dependence, now over the range 
$0\protect\leq r\protect\leq 5$, with $r$ increasing from left to right for
each trace. The efficiency $\protect\eta $ takes on the values 
$\protect\eta =\protect\{0.99,0.90,0.70,0.50\protect\}$ as indicated; in 
all cases, $\protect\bar{n}=0$. Recall that ${\protect\cal B}>2$ heralds a 
direct violation of the CHSH inequality, with the dashed line 
${\protect\cal B}=2$ shown. (b) An expanded view around ${\protect\cal B}=2$. 
Note that ${\protect\cal B}>2$ is impossible for 
$F\protect\leq F_{classical}=\protect\frac{1}{2}$, but that 
${\protect\cal B}>2$ for $F>F_{classical}$ in various domains (including
for $\protect\eta =0.70$ at small $r$). The purported boundary 
$F=\protect\frac{2}{3}$ proposed by Grangier and Grosshans~[11,12] is seen 
to have no particular significance. Contrary to their claims, 
$F=\protect\frac{2}{3}$ provides absolutely no warranty that 
${\protect\cal B}>2$ for $F>\protect\frac{2}{3}$, nor does it preclude 
${\protect\cal B}>2$ for $F<\protect\frac{2}{3}$.}
\label{figure4}
\end{figure}

While the above results follow from the particular form of the CHSH\
inequality introduced by Banaszek and Wodkiewicz\cite{b-k98,b-k99}, we
should note that another quite different path to a demonstration of the
inadequacy of local realism for continous quantum varialbes has recently
been proposed by Ralph, Munro, and Polkinghorne\cite{Ralph00}. These authors
consider a novel scheme involving measurements of quadrature-phase
amplitudes for two entangled beams $(A,B)$. These beams are formed by
combining {\it two} EPR states (i.e., a total of four modes, two for each
beam). Relevant to our discussion is that maximal violations of a CHSH
inequality (i.e., ${\cal B}=2\sqrt{2}$) are predicted for $r\ll 1$, with
then a decreasing maximum value of ${\cal B}$\ for increasing $r$. Once
again, the threshold for onset of the violation of a Bell's inequality
coincides with the threshold for entanglement of the relevant fields [i.e.,
Eq. (\ref{duancondition})], with no apparent significance to the boundary
set by the Heisenberg-type inequality Eq. (\ref{GGqt}) of Grangier and
Grosshans.

To conclude this section, we would like to inject a note of caution
concerning any discussion involving issues of testing Bell's inequalities
and performing quantum teleportation. We have placed them in juxtaposition
here to refute the claims of Grangier and Grosshans related to a possible
connection between the bound $F=\frac{2}{3}$ and violation of Bell's
inequalities (here, via the behavior of the CHSH\ quantity ${\cal B}$ ).
However, in our view there is a conflict between these concepts, with an
illustration of this point provided by the plot of the CHSH quantity ${\cal B%
}$ [Eq.~(\ref{bi})] versus fidelity $F$ [Eq.~(\ref{freta})] in Figure 4. For
example, for $\eta =0.90$, ${\cal B}>2$ over the range $0.50<F\lesssim 0.66$%
, while ${\cal B}<2$ for larger values of $F$. Hence, local hidden variables
theories are excluded for modest values of fidelity $0.50<F\lesssim 0.66$,
but not for larger values $F\gtrsim 0.66$. This leads to the strange
conclusion that quantum resources are required for smaller values of
fidelity but not for larger ones. The point is that the nonseparable states
that can enable quantum teleportation, can {\it in a different context} also
be used to demonstrate a violation of local realism. Again, the
juxtaposition of these concepts in this section is in response to the work
of Ref.\cite{Grangier00a}, which in any event offers no quantitative
evidence in support of their association.

\section{Bell's Inequalities for Scaled Correlations}

The conclusions reached in the preceding section about violations of the
CHSH inequality by the EPR (mixed) state for modes $(1,2)$ follow directly
from the analysis of Banaszek and Wodkiewicz\cite{b-k98,b-k99} as extended
to account for losses in propagation. Towards the end of making these
results more amenable to experimental investigation, recall that the more
traditional versions of the Bell inequalities formulated for spin $\frac{1}{2%
}$ particles or photon polarizations are based upon an analysis of the
expectation value
\begin{equation}
E(\vec{a},\vec{b})  \label{E}
\end{equation}
for detection events at locations $(1,2)$ with analyzer settings along
directions $(\vec{a},\vec{b})$. As emphasized by Clauser and Shimony, actual
experiments do not measure directly $E(\vec{a},\vec{b})$ but rather record a
reduced version due to ``imperfections in the analyzers, detectors, and
state preparation\cite{c-s78}.'' Even after more than thirty years of
experiments, no {\it direct} violation of the CHSH inequality has been
recorded, where by {\it direct} we mean without the need for post-selection
to compensate for propagation and detection efficiencies (also called {\it %
strong violations})\cite{Kwiat94,Fry95}. Rather, only subsets of events that
give rise to coincidences are included for various polarization settings.
This ``problem'' is the so-called detector efficiency loophole that several
groups are actively working to close.

Motivated by these considerations, we point out that an observation of
violation of a Bell-type inequality has recently been reported\cite
{Kuzmich00}, based in large measure upon the earlier proposal of Ref.\cite
{Grangier88}, as well as that of Refs.\cite{b-k98,b-k99}. This experiment
was carried out in a pulsed mode, and utilized a source that generates an
EPR\ state of the form given by Eq.~(\ref{wout}) in the limit $r\ll 1$.
Here, the probability $P(\alpha _{1},\alpha _{2})$ of detecting a
coincidence event between detectors $(D_{1},D_{2})$ for the EPR\ beams $%
(1,2) $ is given by
\begin{equation}
P(\alpha _{1},\alpha _{2})=M[1+V\cos (\phi _{1}-\phi _{2}+\theta )]\text{,}
\end{equation}
with then the correlation function $E$ relevant to the construction of a
CHSH inequality $-2\leq S\leq 2$ given by
\begin{equation}
E(\phi _{1},\phi _{2})=V\cos (\phi _{1}-\phi _{2}+\theta )\text{,}
\end{equation}
where the various quantities are as defined in association with Eqs. (2,3)
in Ref.\cite{Kuzmich00}. Note that the quantity $M$ represents an overall
scaling that incorporates losses in propagation and detection.
Significantly, Kuzmich {\it et al.} demonstrated a violation of a CHSH\
inequality ($S_{\exp }=2.46\pm 0.06$) in the limit $r\ll 1$ and with
inefficient propagation and detection $\eta \ll 1$, albeit with the
so-called ``detection'' or ``fair-sampling'' loophole.

In terms of our current discussion, this experimental violation of a CHSH\
inequality is only just within the nonseparability domain $\Delta
x^{2}+\Delta p^{2}<1$ (by an amount that goes as $\eta r\ll 1$), yet it
generates a large violation of a CHSH\ inequality. If this same EPR state
were employed for the teleportation of coherent states, the fidelity
obtained would likewise be only slightly beyond the quantum-classical
boundary $F_{classical}=\frac{1}{2}$. It would be far from the boundary $F=%
\frac{2}{3}$ offered by Grangier and Grosshans as the point for ``useful
entanglement,'' yet it would nonetheless provide an example of teleportation
with fidelity $F>\frac{1}{2}$ and of a violation of a CHSH inequality. Of
course, the caveat would be the aforementioned ``fair-sampling'' loophole,
but this same restriction accompanies all previous experimental
demonstrations of violations of Bell's inequalities. Once again, we find no
support for the purported significance of the criteria offered by Grangier
and Grosshans\cite{Grangier00a,Grangier00b}.

\section{Conclusions}

Beyond the initial analysis of Ref.\cite{Fuchs00}, we have examined further
the question of the appropriate point of demarcation between the classical
and quantum domains for the teleportation of coherent states. In support of
our previous result that fidelity $F_{classical}=\frac{1}{2}$ represents the
bound attainable by Alice and Bob if they make use only of a classical
channel, we have shown that the nonseparability criteria introduced in Refs.
\cite{duan00,simon00} are sufficient to ensure fidelity beyond this bound
for teleportation with the EPR state of Eq.~(\ref{wout}), which is in
general a mixed state. Significantly, the threshold for entanglement for the
EPR\ beams as quantified by these nonseparability criteria coincides with
the standard boundary between classical and quantum domains employed in
Quantum Optics, namely that the Glauber-Sudarshan phase-space function
becomes non-positive definite\cite{simon00-gs}.

Furthermore, we have investigated possible violations of Bell's inequalities
and have shown that the threshold for the onset of such violations again
corresponds to $F_{classical}=\frac{1}{2}$. For thermal photon number $\bar{n%
}=0$ as appropriate to current experiments, direct violations of a CHSH
inequality are obtained over a large domain in the degree of squeezing $r$
and overall efficiency $\eta $. Significant relative to the claims of
Grangier and Grosshans\cite{Grangier00a,Grangier00b} is that there is a
regime for nonseparability and violation of the CHSH inequality for which $F<%
\frac{2}{3}$ and for which their Heisenberg inequalities are not satisfied.
Moreover, the experiment of Ref.\cite{Kuzmich00} has demonstrated a
violation of the CHSH inequality in this domain for $(r,\eta )\ll 1$ (i.e., $%
F$ would be only slightly beyond $\frac{1}{2}$), albeit with the caveat of
the ``fair-sampling'' loophole. We conclude that fidelity $F>\frac{2}{3}$
offers absolutely no warranty or ``safety'' relative to the issue of
violation of a Bell inequality (as might be desirable, for example, in
quantum cryptography), in direct disagreement with the assertions by
Grangier and Grosshans. Quite the contrary, larger $r$ (and hence larger $F$%
) leads to an exponentially decreasing domain in allowed loss $(1-\eta )$
for violation of the CHSH inequality, as expressed by Eq.~(\ref{bcondition})
\cite{alternate}.

Moreover, beyond the analysis that we have presented here, there are several
other results that support $F_{classical}=\frac{1}{2}$ as being the
appropriate boundary between quantum and classical domains. In particular,
we note that any nonseparable state and hence also our mixed EPR state is
always capable of teleporting perfect entanglement, i.e., one half of a pure
maximally entangled state. This applies also to those nonseparable states
which lead to fidelities $\frac{1}{2}<F\leq \frac{2}{3}$ in coherent-state
teleportation. According to Refs.\cite{Grangier00a,Grangier00b}, this would
force the conclusion that there is entanglement that is capable of
teleporting truly nonclassical features (i.e., entanglement), but which is
not ``useful'' for teleporting rather more classical states such as coherent
states. Further, in Ref.\cite{vanLoock00c} it has been shown that
entanglement swapping can be achieved with two pure EPR states for {\it any
nonzero squeezing} in both initial states. Neither of the initial states has
to exceed a certain amount of squeezing in order to enable successful
entanglement swapping. This is another indication that $F=\frac{2}{3}$,
which is exceedable in coherent-state teleportation only with more than 3 dB
squeezing, is of no particular significance.

We also point out that Giedke et al. have shown that for all bipartite
Gaussian states, nonseparability implies distillability\cite
{Giedke00,Werner00}. This result applies to those nonseparable states for
which $\frac{1}{2}<F\leq \frac{2}{3}$ in coherent state teleportation, which
are otherwise dismissed by Grangier and Grosshans as not ``useful.'' To the
contrary, entanglement distillation could be applied to the mixed EPR states
employed for teleportation in this domain (and in general for $F>\frac{1}{2}$%
)\cite{duan00b}, leading to enhanced teleportation fidelities and to
expanded regions for violations of Bell's inequalities for the distilled
subensemble.

By contrast, there appears to be no support for the claims of Grangier and
Grosshans\cite{Grangier00a,Grangier00b} that their so-called Heisenberg
inequality and information exchange are somehow ``special'' with respect to
the issues of separability and violations of Bell's inequalities. They have
neither found fault in the prior analysis of Ref.~\cite{Fuchs00}, nor with
the application of the work on nonseparability\cite{duan00,simon00,tan99} to
the current problem. They have likewise provided no analysis that directly
supports their assertion that their Heisenberg inequality is in any way
significant to the possibility that ``the behavior of the {\it observed}
quantities can be mimicked by a {\it classical} and {\it local} model.''\cite
{Grangier00a} Rather, they attempt to set aside by {\it fiat\/} a
substantial body of evidence in favor of the boundary $F_{classical}=\frac{1%
}{2}$ for the teleportation of coherent states with a lack of rigor
indicated by their claim that ``$F=\frac{2}{3}$ would be much safer.''\cite
{Grangier00a}

However, having said this, we emphasize that there is no criterion for
quantum teleportation that is sufficient to all tasks. For the special case
of teleportation of coherent states, the boundary between classical and
quantum teleportation is fidelity $F_{classical}=\frac{1}{2}$, as should by
now be firmly established. Fidelity $F>\frac{2}{3}$ will indeed enable
certain tasks to be accomplished that could not otherwise be done with $%
\frac{1}{2}<F\leq \frac{2}{3}$. However, $F=\frac{2}{3}$ is in no sense an
important point of demarcation for entrance into the quantum domain. There
is instead a hierarchy of fidelity thresholds that enable ever more
remarkable tasks to be accomplished via teleportation within the quantum
domain, with no one value being sufficient for all possible purposes. For
example, if the state to be teleported were some intermediate result from a
large-scale quantum computation as for Shor's algorithm, then surely the
relevant fidelity threshold would be well beyond any value currently
accessible to experiment, $F\sim 1-\epsilon $, with $\epsilon \lesssim
10^{-4}$ to be compatible with current work in fault tolerant architectures.
We have never claimed that $F=\frac{1}{2}$ endows special powers for all
tasks such as these, only that it provides an unambiguous point of entry
into the quantum realm for the teleportation of coherent states.

HJK gratefully acknowledges critical input from A. C. Doherty, H. Mabuchi,
E. S. Polzik, and J. P. Preskill, and support from the NSF (Grant No.
PHY-9722674), the Institute for Quantum Information (IQI) funded by the
NSF-ITR Program, and the ONR. SLB and PvL are funded in part under project
QUICOV as part of the IST-FET-QJPC programme. PvL acknowledges support by a
DAAD Doktorandenstipendium (HSP III)


\begin{references}
\bibitem{expvalues}  For the experiment of Ref.\cite{Furusawa98}, the
measured variances were $\Delta x^{2}\approx (0.8\times \frac{1}{2})\approx
\Delta p^{2}$, so that $\Delta x^{2}+\Delta p^{2}\approx 0.8<1$.
\bibitem{notation}  Eq.~(\ref{fscience}) is written in the notation of the
current paper. It follows from Eq.~(2) of Ref.\cite{Furusawa98} for the case
gain $g=1$, Alice's detection efficiency $\eta _{A}^{2}=1$ (denoted $\eta
^{2}$ in \cite{Furusawa98}), and with the replacement of efficiencies $%
(\xi_{1}^{2}=\xi _{2}^{2})$ from \cite{Furusawa98} $\rightarrow \eta $ here.
\bibitem{alphabet}  The expressions of Eqs. (32-35) are strictly applicable
only for the case gain $g=1$ for teleportation of coherent states uniformly
distributed over the entire complex plane. More generally, when working with
a restricted alphabet of states (e.g., coherent amplitudes selected from a
Gaussian distribution), the optimal gain is not unity when referenced to the
fidelity averaged over the input alphabet. In fact as shown in Ref.\cite
{Fuchs00}, the optimal gain is $g=1/(1+\lambda )$ for an input alphabet of
coherent states distributed according to $p(\beta )=(\lambda /\pi
)exp(-\lambda |\beta |^{2})$. When incorporated into the current analysis,
we find that nonseparable EPR states are sufficient to achieve $F>(1+\lambda
)/(2+\lambda )$ (again with optimal gain $g\neq 1$) although $F$ is no
longer a monotonic function of $r$ as in Figure 1. This result is in
complete correspondence with our prior result that $F_{classical}=(1+\lambda
)/(2+\lambda )$ is the bound for teleportation when only a classical channel
is employed. To simplify the discussion, here we set $\lambda =0$
throughout, with then the optimal gain $g=1$ and $F_{classical}=\frac{1}{2}$
.
\bibitem{necandsuff}  Note that for the noisy EPR\ states considered here
with Gaussian statistics, Eqs. (\ref{fidelityduan}) and (\ref{fidelitylimit}%
) are both necessary and sufficient conditions. However, this is not
generally the case for states with Gaussian statistics; there are in fact
nonseparable Gaussian states for which Eq.~(\ref{duancondition}) is not
true. Close examination of Ref.\cite{duan00} shows that their results
concerning necessary and sufficient conditions for states with Gaussian
statistics apply only to Gaussian states in a particular standard form.
Explicit counter examples can be constructed from Ref.\cite{vanLoock00b}.
\bibitem{etatwothirds}  The value $\eta =\frac{2}{3}$ is significant in that
${\cal B}>2$ is not possible for efficiency $\eta \leq \frac{2}{3}$. This
same threshold efficiency was also found by P. H. Eberhard, Phys. Rev. A
{\bf 47}, R747 (1993) in a somewhat different context .
\bibitem{alternate}  We stress that these conclusions relate to the
particular measurement strategy proposed in Refs.\cite{b-k98,b-k99}, which
certainly might be expected not to be optimal in the presence of loss. It is
quite possible that for fixed $\eta $ as $r$ grows, there could be larger
domains and magnitudes for violation of the CHSH inequality if a different
set of observables were chosen, including general POVMs.
\end{references}
\end{document}